\documentclass[aps,pra,reprint,superscriptaddress,nofootinbib]{revtex4-2}

\usepackage{amsmath,amssymb,amsthm,mathtools,bm}
\usepackage{braket}
\usepackage{hyperref}
\usepackage{graphicx}
\usepackage{xcolor}
\usepackage{float}

\theoremstyle{plain}

\theoremstyle{definition}

\theoremstyle{remark}

\usepackage{nag}
\usepackage{graphicx}
\usepackage{amsthm}
\usepackage{tabulary}
\usepackage{threeparttable}
\usepackage{qcircuit}
\usepackage{mathtools}
\usepackage{bm}
\usepackage[table,xcdraw]{xcolor}
\usepackage{datetime}
\usepackage{stackrel}
\usepackage{stmaryrd}
\usepackage{xcolor} 
\usepackage{dsfont}
\usepackage[english]{babel}
\makeatletter
\expandafter\def\csname l@en\endcsname{\csname l@english\endcsname}
\expandafter\def\csname dateen\endcsname{\csname dateenglish\endcsname}
\expandafter\def\csname extrasen\endcsname{\csname extrasenglish\endcsname}
\expandafter\def\csname noextrasen\endcsname{\csname noextrasenglish\endcsname}
\makeatother
\usepackage{units}

\usepackage{tikz}
\usetikzlibrary{backgrounds,decorations.pathreplacing}

\usepackage{chngcntr}
\usepackage[normalem]{ulem}
\usepackage{slashed}
\usepackage{soul}

\usepackage{xcolor}

\usepackage[OT2,T1]{fontenc}
\DeclareSymbolFont{cyrletters}{OT2}{wncyr}{m}{n}
\DeclareMathSymbol{\Sha}{\mathalpha}{cyrletters}{"58}

\usepackage{makeidx}
\makeindex

\usepackage{soul}
\usepackage{xargs}
\newcommandx{\cmnote}[2][1=]{\linespread{1.0}\todo[linecolor=red,backgroundcolor=red!25,bordercolor=red,#1]{#2}}

\let\underline\ul

\allowdisplaybreaks[4]

\let\originalleft\left
\let\originalright\right
\renewcommand{\left}{\mathopen{}\mathclose\bgroup\originalleft}
\renewcommand{\right}{\aftergroup\egroup\originalright}

 \index{} \index{} \index{} \index{} \index{} \index{} \index{} \index{}%
\makeatletter

\newcommand{\ringplus}{\mathbin{\text{\@ringplus}}}

\newcommand{\@ringplus}{%
  \ooalign{\hidewidth\raise1.3ex\hbox{\tiny$\circ$}\hidewidth\cr$\m@th+$\cr}%
}

\newcommand{\ringminus}{\mathbin{\text{\@ringminus}}}

\newcommand{\@ringminus}{%
  \ooalign{\hidewidth\raise0.9ex\hbox{\tiny$\circ$}\hidewidth\cr$\m@th-$\cr}%
}
\makeatother
 \index{} \index{} \index{} \index{} \index{} \index{} \index{} \index{}%

\DeclareFontFamily{U}{wncy}{}
\DeclareFontShape{U}{wncy}{m}{n}{<->wncyr10}{}
\DeclareSymbolFont{mcy}{U}{wncy}{m}{n}
\DeclareMathSymbol{\Sh}{\mathord}{mcy}{"58}

\newcommand{\op}[1]{\hat{#1}}

\renewcommand{\vec}[1]{\bm{#1}}

 \index{} \index{} \index{} \index{} \index{} \index{} \index{} \index{}%
 \index{} \index{} \index{} \index{} \index{} \index{} \index{} \index{}%

 \index{} \index{} \index{} \index{} \index{} \index{} \index{} \index{}%
 \index{} \index{} \index{} \index{} \index{} \index{} \index{} \index{}%
\usepackage{graphicx}

\usepackage{graphicx}
\usepackage{multirow}
\usepackage{hhline}

\xyoption{color}
\xyoption{line}

\newcommandx*\bsbal[3][1=black, 3=->]{\ar @[#1]@{#3} [#2,0] \qw}

\newcommandx*\varbs[4][1=black, 3=\theta, 4=->]{\ar @[#1]@{#4}^{#3} [#2,0] \qw}
\newcommandx*\varbss[4][1=black, 3=\frac{\pi}{4}, 4=->]{\ar @[#1]@{#4}^{#3} [#2,0] \qw}
\newcommandx*\varbsleft[5][1=black, 3=\theta', 4=->]{\ar @[#1]@{#4}^{#3}_{#5} [#2,0] \qw}

\providecommandx*\ctrlg[2]{\control \ar @{-}^{#1} [#2,0] \qw}
\providecommandx*\ctrlog[2]{\controlo \ar @{-}^{#1} [#2,0] \qw}

\makeatletter
 \newcommand{\xmapsfrom}[2][]{%
    \ext@arrow3095\leftarrowfill@{#1}{#2}\mapsfromchar
}
\makeatother

\usepackage{mathtools}

\DeclarePairedDelimiterX\innerp[2]{\langle}{\rangle}{#1\delimsize\vert\mathopen{}#2}%
\DeclarePairedDelimiterX\braketOP[3]{\langle}{\rangle}{#1\,\delimsize\vert\,\mathopen{}#2\,\delimsize\vert\,\mathopen{}#3}%
\DeclarePairedDelimiterX\ketbra[2]{\lvert}{\rvert}{#1\delimsize\rangle\!\delimsize\langle#2}%
\DeclarePairedDelimiterX\outerp[2]{\lvert}{\rvert}{#1\delimsize\rangle\!\delimsize\langle#2}%
\DeclarePairedDelimiterX\projector[1]{\lvert}{\rvert}{#1\delimsize\rangle\!\delimsize\langle#1}%
\begin{document}
%===========================================================

%\title{Finite-squeezing GKP-QPC architectures for photonic quantum memories and one-way repeaters}
\title{Finite-energy GKP-QPC architectures for photonic quantum memories and repeaters}
\author{Kaustav Chatterjee}
 %\altaffiliation[Also at ]{Physics Department, XYZ University.}%Lines break automatically or can be forced with \\
 \email{kauch@dtu.dk}
\author{Ulrik Lund Andersen}%
 \email{ulrik.andersen@fysik.dtu.dk}
\affiliation{%
Center for Macroscopic Quantum States (bigQ), Department of Physics, Technical University of Denmark, \\
 Building 307, Fysikvej, 2800 Kongens Lyngby, Denmark %\textbackslash\textbackslash
}%
\date{\today}

\begin{abstract}
Photonic quantum networks require error-correction architectures that remain useful with finite-energy bosonic states, pure-loss fiber transmission, and can explicitly account for resources. In this light, we study a concatenated architecture in which each physical rail is a finitely squeezed Gottesman--Kitaev--Preskill (GKP) qubit transmitted through a pure-loss fiber segment, corrected by teleportation-based GKP error correction with finitely squeezed ancillae, and decoded by an outer quantum parity code (QPC). The GKP layer converts continuous homodyne syndromes into effective rail-level Pauli marginals, while the QPC layer suppresses the residual qubit-level errors. This gives a common finite-squeezing framework for all-optical quantum memories and one-way repeaters without pre-amplification. For the concatenated code family considered here, we find a finite-squeezing threshold of $5.06\,\mathrm{dB}$ at zero propagation loss. In the memory setting, the QPC layer lowers the squeezing at which repeated error correction becomes beneficial from $6.7\,\mathrm{dB}$ for bare GKP correction to $5.2\,\mathrm{dB}$ for QPC$(3,3)$ and $4.3\,\mathrm{dB}$ for QPC$(5,5)$, and improves the average-fidelity ratio by up to $75$--$90\%$ in the relevant intermediate-noise regime. In the repeater setting, avoiding pre-amplification gives larger secret-key fractions at moderate squeezing along with production of an optima beyond which the key rates degrade. Resource-normalized rates show that QPC concatenation can exceed the repeaterless PLOB benchmark by orders of magnitude and extend the communication reach, at short repeater spacing, to distances of order $10^4$km with $14$dB squeezing. 
%However, QPC concatenation becomes detrimental when each elementary hop is too lossy. Optimizing over QPC size reveals interior optima, rather than a monotonic preference for larger codes. These results provide quantitative design rules for finite-squeezing GKP--QPC quantum memories and repeaters.
\end{abstract}

\maketitle

%===========================================================
\section{Introduction}

Quantum cryptographic and quantum computing tasks offer qualitative advantages over their classical counterparts~\cite{huang2025vastworldquantumadvantage,grover1996fastquantummechanicalalgorithm,Shor_1997}. Realizing these advantages over large distances requires quantum networks that can distribute or transmit quantum information despite photon loss and operational noise. This challenge has motivated a broad range of quantum-repeater architectures~\cite{bigelRevLett.81.5932,Munro2015InsideQR,RevModPhys.95.045006}. A useful classification divides repeaters into three generations~\cite{Muralidharan2015OptimalAF}. The first two rely on heralded entanglement generation and purification, and therefore require quantum memories to store states while waiting for successful heralding events. Third-generation repeaters instead use one-way quantum error correction at each repeater station. This removes the need for long-lived heralding memories and can yield higher rates, but shifts the central challenge to the quality, experimental implementation, and resource overhead of the underlying error-correcting code.

Here we consider a photonic implementation based on a concatenation of a bosonic inner code and a discrete-variable outer code. Each physical rail is encoded using the continuous-variable Gottesman-Kitaev-Preskill (GKP) code~\cite{Gottesman_2001}, while the outer layer is a quantum parity code (QPC)~\cite{Ralph_2005}. These two codes play complementary roles. The GKP code is naturally suited to optical modes because continuous homodyne syndromes can be used to infer small displacement errors. After GKP error correction, the remaining noise can be represented as effective rail-level qubit errors. The QPC then suppresses these residual discrete errors through redundancy across multiple rails. The central idea of this work is therefore to use the GKP layer to convert continuous-variable loss noise into effective rail-level Pauli error statistics, and then use the QPC layer to suppress the remaining logical errors.

%Here we focus on photonic implementations, where optical modes are naturally compatible with room-temperature operation and telecom transmission. We encode each rail using the continuous-variable Gottesman--Kitaev--Preskill (GKP) code~\cite{Gottesman_2001}, and concatenate this inner bosonic code with a discrete-variable quantum parity code (QPC)~\cite{Ralph_2005}. 

\begin{figure}[t]
    \centering
    \includegraphics[width=0.95\linewidth]{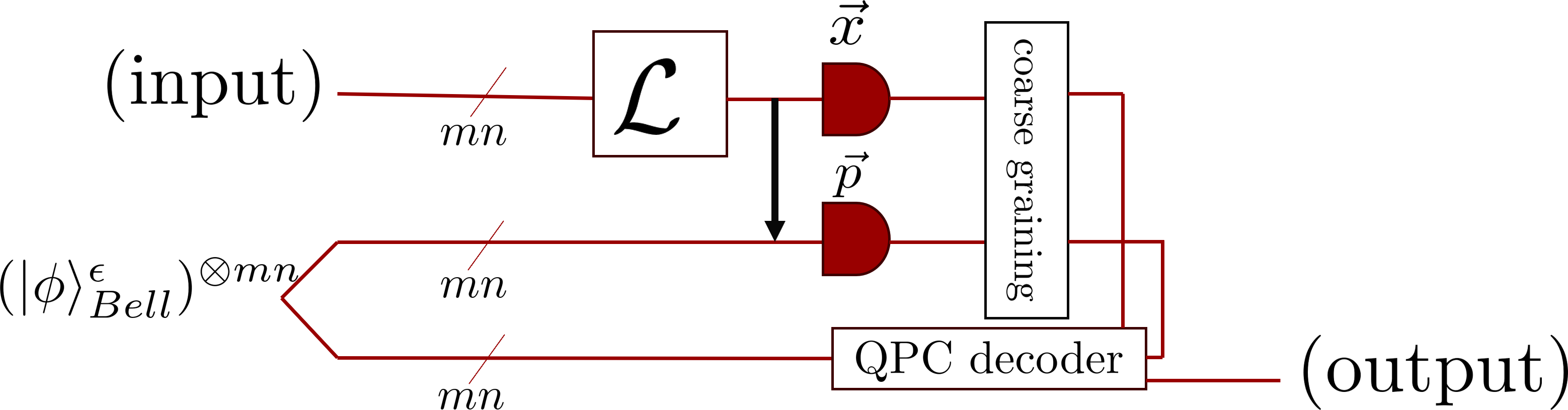}
    \caption{
    Conceptual structure of the finite-squeezing GKP-QPC architecture. Each physical rail is a finitely squeezed GKP qubit (input) transmitted through a pure-loss fiber segment and corrected by teleportation-based GKP error correction using finitely squeezed GKP Bell resources. The $mn$ rails represent the physical redundancy of the outer QPC$(n,m)$ code.  The continuous homodyne syndromes are converted into effective rail-level Pauli error marginals, which are then decoded by an outer QPC. The same elementary module can be repeated in time to form an all-optical quantum memory or concatenated in space to form a one-way quantum repeater. 
    }
    \label{fig:intro_architecture}
\end{figure}

The elementary module studied in this work is illustrated schematically in Fig.~\ref{fig:intro_architecture}. A finitely squeezed GKP qubit is transmitted through a pure-loss fiber segment and then corrected using teleportation-based GKP error correction with finitely squeezed qunaught ancillae. The resulting homodyne outcomes are coarse-grained into effective correct/incorrect probabilities in the relevant $X$- and $Z$-error sectors. These rail-level error marginals are passed to the QPC decoder, which produces a logical Pauli-frame update. This gives a single finite-squeezing framework for both all-optical quantum memories and one-way quantum repeaters.

Closely related GKP--QPC concatenations have been studied previously~\cite{Muralidharan_2014,Fukui_2023}, but mostly in code-capacity or additive-Gaussian-noise settings with ideal GKP states or ideal ancillae. Our aim is to benchmark a more physical model: All GKP states and teleportation ancillae have finite squeezing, the transmission channel is treated as pure loss, and the outer-code performance is reported per occupied optical mode. We also deliberately avoid pre-amplification. Pre-amplifying a lossy GKP state can convert pure loss into an effective random-displacement channel~\cite{Noh_2019}, but this can obscure the finite-energy peak-shifting effect that is intrinsic to pure loss. At low GKP squeezing, the dominant error mechanism is the broadening of the GKP peaks. At high squeezing, however, the peaks become narrower but also more sensitive to loss-induced inward displacement in phase space. As a result, increasing the GKP squeezing is not always beneficial. This competition between peak broadening and peak shifting leads to an optimal squeezing, which becomes an important design parameter for the repeater architecture. 
%we show avoiding pre-amplification has a benifit in one way repeater scenarios as it allows for comparatively higher secret key rates at lower GKP squeezing which itself is an experimental bottleneck.

We apply the same GKP--QPC module in two settings. First, we consider an all-optical quantum memory in which the logical state is repeatedly transmitted through short fiber segments and corrected, following the spirit of Ref.~\cite{chatterjee2026allopticalquantummemoryusing}. In this setting, the central question is whether the QPC layer can make repeated GKP correction beneficial at squeezing values where the bare GKP memory is still below threshold. 
%We find that it can: the onset of useful repeated correction is shifted to lower squeezing, and the relative fidelity improvement is largest in the intermediate regime where the GKP layer is good enough to leave mostly correctable rail errors but not so good that the outer code is redundant.
Second, we consider a one-way repeater chain in which the same correction module is placed at each repeater station. In this case, we compare bare GKP repeaters with GKP--QPC repeaters, normalize the secret-key fraction by the number of occupied optical modes, and benchmark the performance against the repeaterless PLOB bound~\cite{Pirandola_2017}. 

Our main findings are as follows. At zero propagation loss, where finite GKP ancilla squeezing is the only noise source, we find a finite-squeezing threshold of $5.06\,\mathrm{dB}$ for the concatenated GKP--QPC architecture. In the memory setting, the QPC layer lowers the squeezing at which repeated correction becomes useful from $6.7\,\mathrm{dB}$ for bare GKP correction to $5.2\,\mathrm{dB}$ for QPC$(3,3)$ and $4.3\,\mathrm{dB}$ for QPC$(5,5)$. The relative fidelity improvement is largest in the intermediate-noise regime, where the GKP layer is sufficiently accurate to leave mostly correctable rail errors but not so accurate that the outer code becomes redundant.

%The resulting picture is not simply that more redundancy is always better. QPC concatenation improves the reach for short repeater spacing and moderate-to-high squeezing, but it becomes harmful when the segment length is too large and each physical rail is already too noisy. We therefore identify both an optimal squeezing, caused by the competition between peak broadening and peak shifting, and an optimal QPC size, caused by the competition between logical error suppression and optical-mode overhead. 

In the repeater setting, avoiding pre-amplification gives higher secret-key fractions at moderate squeezing, which is experimentally relevant because high-quality GKP squeezing remains a major bottleneck. The advantage is not monotonic: at high squeezing, loss-induced peak shifting reduces performance and produces an optimal squeezing. With the outer QPC included, the repeater can exceed the repeaterless PLOB bound by orders of magnitude in the low-loss segment regime. However, QPC concatenation is not universally beneficial. It improves the reach for short repeater spacing and moderate-to-high squeezing, but becomes detrimental when the elementary segment is too lossy and each physical rail is already outside the useful operating regime of the outer code. Optimizing over the QPC parameters therefore reveals interior optima rather than a monotonic preference for larger codes.

The paper is organized as follows. Section~\ref{sec:a1} reviews the finite-squeezing GKP error-correction model. Section~\ref{sec:a2} derives the concatenated GKP--QPC effective channel. Section~\ref{sec:a3} applies it to optical-fiber memories. Section~\ref{sec:architecture_rates} applies the same framework to one-way repeaters.

\section{Preliminaries}\label{sec:a1}

\subsection{GKP code}
We start with a brief review of the GKP code~\cite{Gottesman_2001,Brady_2024}. Specifically, we consider the square GKP code which encodes a qubit into a single-mode continuous-variable (CV) Hilbert space $\mathcal{H}$, described by position ($\hat{q}$) and momentum ($\hat{p}$) operators satisfying $[\hat{q},\hat{p}]=i$. The ideal GKP codespace is the simultaneous $(+1)$-eigenspace of the two commuting stabilizers
\begin{equation}
\hat S_1 = e^{-2 i \sqrt{\pi}\,\hat p}, 
\qquad
\hat S_2 = e^{ 2 i \sqrt{\pi}\,\hat q}
\end{equation}
and is spanned by the ideal codewords
\begin{equation}
|0\rangle = \sum_{s\in\mathbb Z} |(2s)\sqrt{\pi}\rangle_q,
\qquad
|1\rangle = \sum_{s\in\mathbb Z} |(2s+1)\sqrt{\pi}\rangle_q,
\end{equation}
where $|q_0\rangle_q$ denotes the $\hat q$-eigenstate with eigenvalue $q_0$.
These ideal codewords are unphysical because they possess infinite energy. Physically valid states are constructed by applying the damping operator
\begin{equation}
\hat N(\varepsilon) = e^{-\varepsilon \hat n},
\qquad
\hat n=\tfrac12(\hat q^2+\hat p^2-1),
\end{equation}
where the parameter $\varepsilon$ is related to the usual squeezing value in dB via $-10\log_{10}(\tanh \varepsilon)$~\cite{PhysRevA.108.052413,Tzitrin_2020}. The approximate GKP codewords are then typically given by
\begin{equation}
|\bar 0^\varepsilon\rangle \propto \hat N(\varepsilon)\,|0\rangle,
\qquad
|\bar 1^\varepsilon\rangle \propto \hat N(\varepsilon)\,|1\rangle
\end{equation}
along with the proper normalisation $\langle \bar 0^\varepsilon|\bar 0^\varepsilon\rangle=\langle \bar 1^\varepsilon|\bar 1^\varepsilon\rangle=1$. Following the Bloch-vector representation for the GKP subspace introduced in~\cite{PhysRevA.108.052413}, any non-ideal GKP state can be written as a four-component Bloch vector $\vec{a}:=(a_0,a_1,a_2,a_3)$, corresponding to the state
\begin{equation}
    \rho^\varepsilon=\sum_i a_i\sigma^\varepsilon_i
\end{equation}
where $\sigma^\varepsilon_i$ is the damped version of the ideal GKP Pauli operator, i.e. $\sigma^\varepsilon_i=\hat{N}(\varepsilon)\sigma_i \hat N(\varepsilon)$.

\subsection{Teleportation-based GKP error correction}
For our quantum memory and repeater schemes, we consider GKP states traveling through optical fibers, which can be modeled as a Gaussian pure-loss channel $\mathcal{L}_\eta$ with Stinespring dilation
\begin{equation}
\mathcal{L}_\eta(\hat \rho_S)=\mathrm{tr}_E\!\Big(\op{B}_\eta(\hat \rho_S\otimes \ket{\mathrm{vac}}\!\bra{\mathrm{vac}}_E)\op{B}_\eta^\dagger\Big),
\end{equation}
where $\op{B}_\eta$ denotes a beam-splitter unitary coupling the system mode ($S$) to an environmental vacuum mode ($E$),
\begin{equation}
\begin{split} \label{cir:beamsplitter}
	\raisebox{-1.2em}{$\op{B}_{\eta}  =
 e^{ - i\theta ( \op{q}_S \op{p}_E - \op{p}_S \op{q}_E )} \, = \, $}
         \Qcircuit @C=1.25em @R=2.5em @! 
         {
          &  \varbs{1} & \rstick{S}  \qw \\
          & \qw       & \rstick{E} \qw
  		  } 
\end{split}\quad\quad,
\end{equation}
The transmissivity of the beam splitter is defined by $\cos^2(\theta)=\eta$. The effect of loss on an approximate GKP state is twofold. First, the variance of the Gaussian peaks increases due to added vacuum noise. Second, the individual peaks are displaced toward the origin relative to their initial positions in phase space. Both effects lead to logical errors by pushing probability distributions into incorrect quadrature bins. However, the relative importance of these two effects depends strongly on the squeezing level of the GKP state. For low squeezing, the increased peak variance is the dominant source of logical errors, whereas, in the high-squeezing regime, the dominant mechanism is the inward displacement of peaks located far from the origin. Ref.~\cite{PhysRevA.108.052413} showed that this leads to an optimal squeezing value $\varepsilon$ for a given transmissivity $\eta$. To manage the CV noise introduced by photon loss, we use teleportation-based GKP error correction~\cite{PhysRevA.102.062411}.
The key resource of this protocol is the GKP qunaught state, defined as
\begin{align}
    \ket{\varnothing}=\sum_{s\in \mathbb{Z}}\ket{s\sqrt{2\pi}}_q.
\end{align}
When two such states are mixed on a balanced beam-splitter, they create an approimate GKP Bell pair, $\hat B_{\frac{\pi}{4}}\ket{\varnothing\varnothing}\propto\ket{00}+\ket{11}$ which we denote by $\ket{\phi}^\epsilon_{Bell}$. Together with a Bell measurement consisting of another balanced beam-splitter and two homodyne measurements, this resource enables the teleportation-based error-correction circuit:
\begin{equation} 
\begin{split} \label{Cir:teleport}
    \Qcircuit @C=2.0em @R=2.5em  
    {
    &&&\lstick{\text{(in)}}  &\qw & \varbss{1} &  \qw&\ket{q_0}_q \\
    &&&\lstick{\ket{\varnothing^\varepsilon}}  & \varbss{1} &  \qw & \qw&\ket{p_0}_p\\
    &&&\lstick{\ket{\varnothing^\varepsilon}}  & \qw &  \qw & \qw & (\text{out})
    }
\end{split}
\qquad\qquad
\end{equation}
Here, the approximate qunaught states are given by $\ket{\varnothing^\varepsilon}\propto\hat N(\varepsilon)\ket{\varnothing}$ with $\langle \varnothing^\varepsilon|\varnothing^\varepsilon\rangle=1$.
Operationally, teleportation-based GKP error correction projects the lossy state back into the logical GKP codespace and converts CV noise into discrete logical Pauli errors. For a given syndrome outcome $(q_0,p_0)$ we determine the corresponding logical Pauli-frame update $I,X,Y$, or $Z$ for the output state of the circuit in Eq.~\eqref{Cir:teleport}. This requires a decoding strategy that maps the CV syndromes to logical operations. For the remainder of this work, we adopt the decoding strategy of~\cite{chatterjee2026allopticalquantummemoryusing}. The entire teleportation-based error correction scheme can be written as a map
\begin{equation}
    \mathcal{E}_t(\rho)=p_I\rho+p_X X\rho X+p_Y Y\rho Y+p_Z Z\rho Z
\end{equation}
Here, $p_X$ is the hard-decision probability of a logical $X$ error, obtained by integrating over all syndromes that yield $X$ as the logical error. Analogous definitions apply to $p_Y$ and $p_Z$, while $p_I$ is the probability of no logical Pauli error. Expressed in terms of the Bloch-vector transformation, one can show~\cite{chatterjee2026allopticalquantummemoryusing} that one round of photon loss, GKP error correction, and logical Pauli-frame updating is described by the effective transfer matrix $C^{\varepsilon, \eta}$.

\section{Error correction for concatenated architecture}\label{sec:a2}

We now describe how the finite-squeezing GKP recovery channel is converted into an effective logical QPC channel. We use the convention QPC$(n,m)$: each of the $n$ blocks contains $m$ physical qubits. At the block level,
\begin{align}
    |0\rangle^{(m)} &= |0\rangle^{\otimes m}, \\
    |1\rangle^{(m)} &= |1\rangle^{\otimes m}.
\end{align}
The logical $X$-basis states are
\begin{equation}
    |\pm\rangle_L = \left( \frac{|0\rangle^{\otimes m} \pm |1\rangle^{\otimes m}}{\sqrt{2}} \right)^{\otimes n} .
\end{equation}
The physical qubits in these blocks are GKP-encoded optical modes. The QPC itself is the loss-tolerant photonic code introduced in Ref.~\cite{Ralph_2005}, while QPC decoding for bosonic concatenation was analyzed in detail in the ideal-GKP, random-displacement noise setting of Ref.~\cite{Fukui_2023}. Here we adapt the same outer-code decoding logic to a more physical setting: replacing the circuit of \cite{Fukui_2023} by macronode lattice \cite{PhysRevA.102.062411} along with assuming finitely damped GKP states transmitted through pure loss channel.

\subsection{From the single-rail GKP map to the logical QPC map}

\subsubsection{Single-shot circuit}

\begin{figure}[H]
    \centering
    \includegraphics[width=1\linewidth]{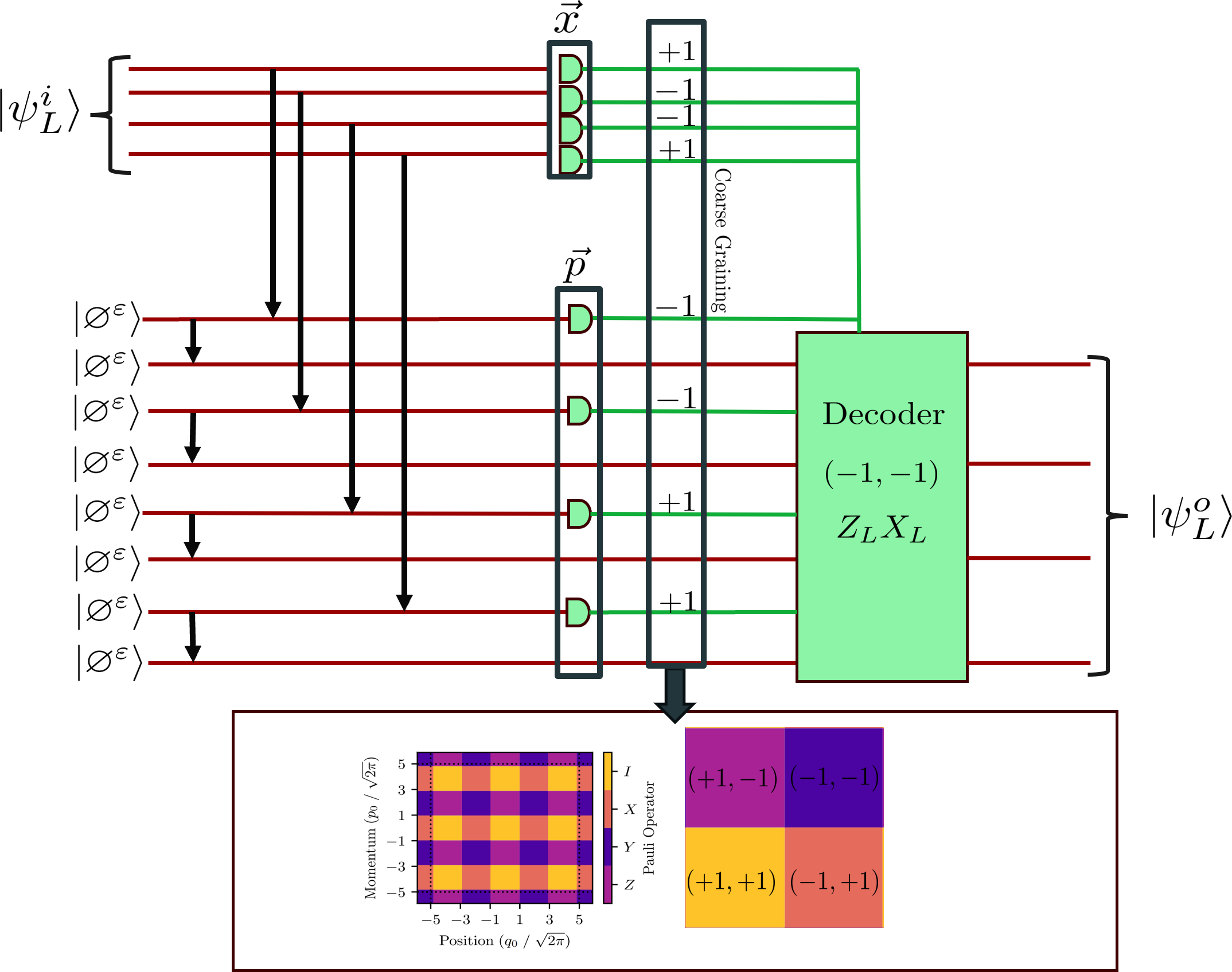}
    \caption{Circuit diagram for a single round of error correction with QPC$(2,2)$, generalizable to QPC$(n,m)$. Each physical rail undergoes teleportation-based GKP error correction. The corresponding homodyne outcome is classically coarse-grained into sector signs, which are fed into the outer QPC decoder~\cite{Fukui_2023} (also see appendix \ref{app:qpc_hard_decoder}). The resulting logical Pauli-frame update can either be tracked classically or implemented by the corresponding logical feed-forward displacement. The arrows are 50-50 beam-splitters.}
    \label{fig:betax}
\end{figure}

A representative single round is shown in Fig.~\ref{fig:betax} for QPC$(2,2)$. For QPC$(n,m)$ the round uses $2mn$ approximate qunaught ancillae $|\varnothing^\varepsilon\rangle$, two for each physical rail. Apart from homodyne measurements and classical feed-forward, the optical processing is Gaussian. The single-rail GKP decoder maps each homodyne outcome $(q_0,p_0)$ to a Pauli-frame update. We record this update as two signs,
\begin{equation}
\begin{split}
    &I\to(+1,+1),\quad
    Z\to (+1,-1),\quad\\
    &X\to (-1,+1),\quad
    Y\to(-1,-1),
\end{split}
\end{equation}
where the two entries correspond to the $Z$- and $X$-sector error information used by the outer QPC decoder. These signs are then processed by the QPC majority/parity decoding rules of Ref.~\cite{Fukui_2023}.

\subsubsection{Bloch-vector reduction}

Let $C^{\varepsilon,\eta}$ be the single-segment GKP Pauli-transfer matrix (PTM) obtained from the finite-squeezing simulation of Ref.~\cite{chatterjee2026allopticalquantummemoryusing}. The QPC decoder of Ref.~\cite{Fukui_2023} does not need the full physical Pauli distribution in every step. For each basis $B\in\{X,Z\}$ it needs the probability that the rail-level outcome is correct or incorrect in that basis. In the present work we use deterministic hard-decision GKP recovery, so no rail is discarded; equivalently, the erasure probability is set to zero. This keeps the comparison focused on finite squeezing and pure loss. Erasure-aware decoding can be incorporated by retaining a third outcome in the same QPC decoder, but it is not used in the numerical results below primarily because we find that allowing for erasures does not severely change the threshold listed below. The exact analysis of the decoder and the way its implemented in code is collected in the appendix \ref{app:qpc_hard_decoder}. 

The required sector probabilities are extracted by probing the single-rail PTM with the ideal $X$- and $Z$-eigenstate Bloch vectors,
\begin{equation}
    v_X = (1,1,0,0)^T,
    \qquad
    v_Z = (1,0,0,1)^T .
\end{equation}
Define
\begin{equation}
    \ell_X := C^{\varepsilon,\eta} v_X,
    \qquad
    \ell_Z := C^{\varepsilon,\eta} v_Z,
\end{equation}
and normalize by the output trace components,
\begin{equation}
    r_X := \frac{(\ell_X)_1}{(\ell_X)_0},
    \qquad
    r_Z := \frac{(\ell_Z)_3}{(\ell_Z)_0}.
\end{equation}
The corresponding single-rail correct/incorrect probabilities are
\begin{align}
    P_c^{(X)} &= \frac{1+r_X}{2},
    &
    P_i^{(X)} &= \frac{1-r_X}{2},
    \\
    P_c^{(Z)} &= \frac{1+r_Z}{2},
    &
    P_i^{(Z)} &= \frac{1-r_Z}{2}.
\end{align}
For a diagonal physical Pauli channel
\begin{equation}
    \mathcal{E}_t(\rho)=p_I\rho+p_X X\rho X+p_Y Y\rho Y+p_Z Z\rho Z,
\end{equation}
these reduce to
\begin{align}\label{corrincorr}
    P_i^{(X)} &= p_Z + p_Y,
    &
    P_c^{(X)} &= p_I + p_X,
    \\
    P_i^{(Z)} &= p_X + p_Y,
    &
    P_c^{(Z)} &= p_I + p_Z .
\end{align}
Thus the outer decoder consumes the two sector marginals rather than the full four-component Pauli distribution. Feeding these marginals into the QPC majority/parity decoder gives two logical sector-error probabilities, denoted $E_X$ and $E_Z$. Because the QPC decoder processes $Z$-type and $X$-type errors through independent parity and majority-vote chains, the logical $X$ and $Z$ errors are statistically independent conditioned on the i.i.d.\ physical error channel. We then construct the logical channel by using the sector-independent Pauli distribution with these marginals,
\begin{align}
    p_I^{L} &= (1-E_X)(1-E_Z), \\
    p_Z^{L} &= E_X(1-E_Z), \\
    p_X^{L} &= (1-E_X)E_Z, \\
    p_Y^{L} &= E_XE_Z .
\end{align}
Equivalently,
\begin{equation}
    \mathcal{E}_{t_0}^L(\rho_L)
    := p_I^{L}\rho_L+p_X^{L}X\rho_LX+p_Y^L Y\rho_LY+p_Z^L Z\rho_LZ .
\end{equation}
where $t_0$ takes care of unit memory time per round given in terms of transmissivity ($\eta(L)=e^{-L/L_{att}}$) as $t_0=l/c_{{fib}}$ with the attenuation length $L_{att}=22\,\mathrm{km}$~\cite{hler2024quantumrepeatersbasedstationary}. . The corresponding logical PTM is
\begin{equation}\label{large}
C^{\varepsilon,\eta}_L=\mathrm{diag}\bigl(1,\lambda_X,\lambda_Y,\lambda_Z\bigr),
\end{equation}
with
\begin{align}
    \lambda_X &= 1 - 2E_X, \\
    \lambda_Z &= 1 - 2E_Z, \\
    \lambda_Y &= (1-2E_X)(1-2E_Z).
\end{align}
This PTM acts on the logical QPC Bloch vector after one lossy segment and one complete GKP--QPC correction round. The modeling assumptions used throughout the numerics are therefore: independent and identically distributed physical rails, deterministic hard-decision GKP recovery, sector-marginal QPC decoding, and no additional circuit noise from switching, homodyne inefficiency, or imperfect Gaussian gates. 
%It is also worth mentioning that performing Monte-carlo simulation using the single shot circuit converges over a large number of shots to the Bloch-vector reduction.
As a validation, we compared the Bloch-vector reduction with direct Monte Carlo simulations of the single-shot circuit for representative values of squeezing and loss. The decoded logical sector-error probabilities agreed within $O(10^{-3})$ level at 5 million Monte-carlo shots, confirming that the reduced description captures the relevant QPC decoder output along with the fact that the minor discrepancy comes from discarding of correlations between errors in our model.

\begin{figure}[h]
    \centering
    \includegraphics[width=1\linewidth]{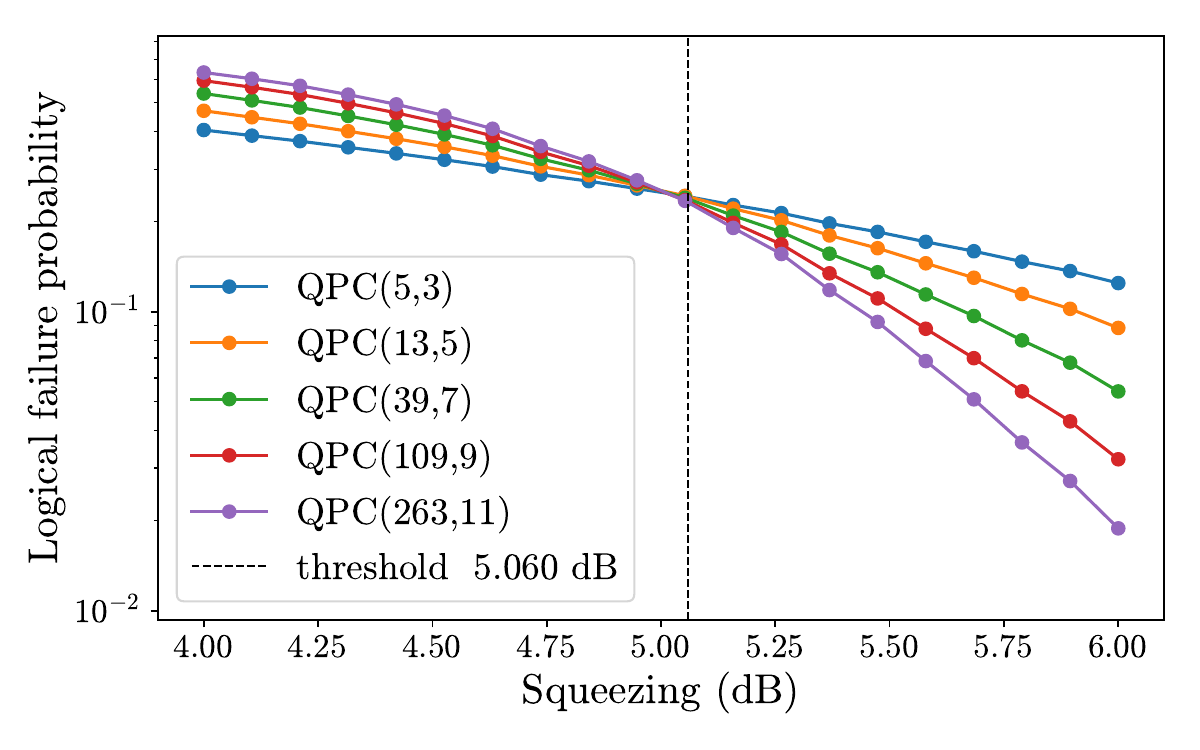}
    \caption{Logical failure probability for the GKP--QPC$(n,m)$ concatenated code at zero propagation loss ($\eta=1$), as a function of GKP squeezing. For each repetition number $m$ the block size $n$ is optimized to minimize the failure probability. The crossing of the curves identifies the finite-squeezing threshold at $5.06\,\mathrm{dB}$.}
    \label{fig:threshold}
\end{figure}

Figure~\ref{fig:threshold} gives the resulting finite-squeezing threshold at zero propagation loss, $\eta=1$, so that the only noise source is the finite squeezing of the GKP ancillae used in teleportation-based correction. The plotted quantity is the logical failure probability $1-p_I^L$. The curves cross at $5.06\,\mathrm{dB}$. This is higher than the $2.01\,\mathrm{dB}$ code-capacity threshold of Ref.~\cite{Fukui_2023}, as expected: finite-squeezing ancillae add recovery noise at every rail before the outer QPC can suppress the resulting logical errors. 
We define the finite-squeezing threshold as the squeezing value at which increasing the QPC size changes from increasing to decreasing the logical failure probability, with the block size $n$ optimized for each repetition number $m$. The value $5.06$ dB is therefore a numerical threshold for this code family, finite-squeezing recovery model, and decoder, rather than a universal threshold for all possible GKP--QPC concatenations. Further, allowing for erasures and a subsequent optimization over them cause the threshold to lower by $0.07$ dB.

\section{Architecture for quantum memory}\label{sec:a3}

The quantum memory architecture considered here follows the same basic principle as the all-optical memory proposed in~\cite{chatterjee2026allopticalquantummemoryusing}. A logical state is repeatedly transmitted through lossy fiber segments and corrected after each segment. The central question is how well the logical state is preserved when every physical rail of the QPC code is subjected to pure loss followed by the concatenated GKP--QPC error-correction procedure. 
 composed over many rounds over which the state stays within the memory. The figure of merit is the average fidelity of the resulting logical error-correction channel, which is described by $\mathcal{E}^L_{t_s}$. Since this is an effective qubit channel at the logical level, its average fidelity can be written in closed form in terms of the corresponding PTM~\cite{Nielsen_2002}:
\begin{equation}\label{eq:avgBloch}
    F_\text{avg}(\mathcal E^L_{t_s})=\frac{1}{12}\sum_{\vec a_i\in\mathcal{B}}\vec a_{i}\cdot ( C^{\varepsilon,\eta}_L)^{N_{\mathrm{EC}}}
    (\vec a_{i}).
    \end{equation}
With the corresponding Bloch vector set $\mathcal B=\{(1,\pm 1,0,0),(1,0,\pm 1,0),(1,0,0,\pm 1)\}$, corresponding to the six cardinal pure states of a qubit in the Bloch representation.

With a speed of light in fibre $c_{fib}=2.07\times 10^8\,\mathrm{m/s}$,corresponding to $n_\text{fib}= 1.45$, the storage time is given by
\begin{align}\label{eq:storagetime}
    t_s= N_{\mathrm{EC}}t_0=Nl/c_{fib},
\end{align}
%while the loss per fibre segment of length $L$ is described by the transmissivity
%\begin{align}
%    \eta(L) = \exp\!\left(-\frac{L}{L_{\mathrm{att}}}\right)
%\end{align}
%with the attenuation length $L_{\mathrm{att}}=22\,\mathrm{km}$~\cite{hler2024quantumrepeatersbasedstationary}. 
Here $N_{\mathrm{EC}}$ is the number of error-correction rounds over which the logical map $\mathcal{E}^L_{t_0}$ is applied. 

The physical qubits of the QPC$(n,m)$ code are finitely squeezed GKP-encoded optical modes. In each memory round, every rail first undergoes the pure-loss channel $\mathcal{L}_\eta$ and then enters the teleportation-based GKP error-correction circuit. The resulting syndrome information $(q_0,p_0)$ is processed using the single-rail GKP decoder of~\cite{chatterjee2026allopticalquantummemoryusing}, yielding the effective rail-level error probabilities $(p_X,p_Y,p_Z)$. This coarse-grained per-rail information is then passed to the QPC$(n,m)$ decoder, which performs classical post-processing and determines the appropriate logical Pauli-frame update.

After correction, the logical state is routed back into the memory loop and subjected to another lossy segment followed by another correction round. This process is repeated for $N_\mathrm{EC}$ rounds before the logical state is finally read out. Throughout the present analysis, we neglect switching loss as well as imperfections in beam splitters, homodyne measurements, and other Gaussian operations. The results therefore isolate the effect of finite GKP squeezing, fiber loss, and the outer QPC encoding. One additional implementation detail is that the damped-ideal operator basis $\hat N(\varepsilon)\sigma_j\hat N(\varepsilon)$,is generally nonorthogonal. Before interpreting its components as qubit Bloch coordinates, we transform the map to the orthonormal finite-energy logical basis $\{|+^\varepsilon\rangle,|-^\varepsilon\rangle\}$. Let
$S_\varepsilon$ denote the corresponding change-of-basis matrix. The logical-qubit PTM used
below is therefore
\[
C_{L,\mathrm{orth}}^{\varepsilon,\eta}
   =S_\varepsilon^{-1}C^{\varepsilon,\eta}_LS_\varepsilon .
\]
For notational simplicity, we suppress the subscript ``orth'' in the
remainder and write this matrix as $C^{\varepsilon,\eta}_L$. Consequently, its
zeroth output component gives the output trace, while the normalized first
and third components give the physical $X$- and $Z$-Bloch coordinates,
respectively.

%Noise due to switching as well as imperfections in beam-splitters and homodyne measurements is neglected in the present analysis. The physical qubits of the QPC$(n,m)$ code are constructed from finitely squeezed GKP states; each rail undergoes the loss channel $\mathcal{L}_\eta$ before entering the teleportation-based GKP error correction circuit. The resulting soft syndrome information $(q_0,p_0)$ is averaged using the internal single-rail GKP decoder of~\cite{chatterjee2026allopticalquantummemoryusing} to yield the marginals $(p_X,p_Y,p_Z)$. This coarse-grained per-rail information is then fed to the QPC$(n,m)$ decoder, which performs classical post-processing to determine the appropriate logical operation. The corrected logical state is switched back to the input, and the process is repeated for $n$ rounds before the logical state is read out.

\subsection{Optimal segment length}
Given a GKP squeezing $\varepsilon$ and a total fiber length $L_{\mathrm{tot}}$, one can perform concatenated error correction $N_{\mathrm{EC}}$  times, with each correction applied after a segment of length $L_\mathrm{seg}=L_{\mathrm{tot}}/N_{\mathrm{EC}}$. Optimizing over segment lengths yields the optimal value $L_\mathrm{opt}=L_{\mathrm{tot}}/N_{\mathrm{EC}}^\mathrm{opt}$ for a given $\varepsilon$. Physically, this optimization balances two competing effects. If the segment length is too long, photon loss accumulates before correction is applied. If the segment length is too short, the finite-squeezing noise introduced by each teleportation-based correction step becomes dominant. This optimization was performed in~\cite{chatterjee2026allopticalquantummemoryusing} for the bare GKP case, where a squeezing threshold of approximately $6.7\,\mathrm{dB}$ was identified below which the optimal segment length diverges and error correction becomes counterproductive. 

\begin{figure}[h]
    \centering
    \includegraphics[width=1\linewidth]{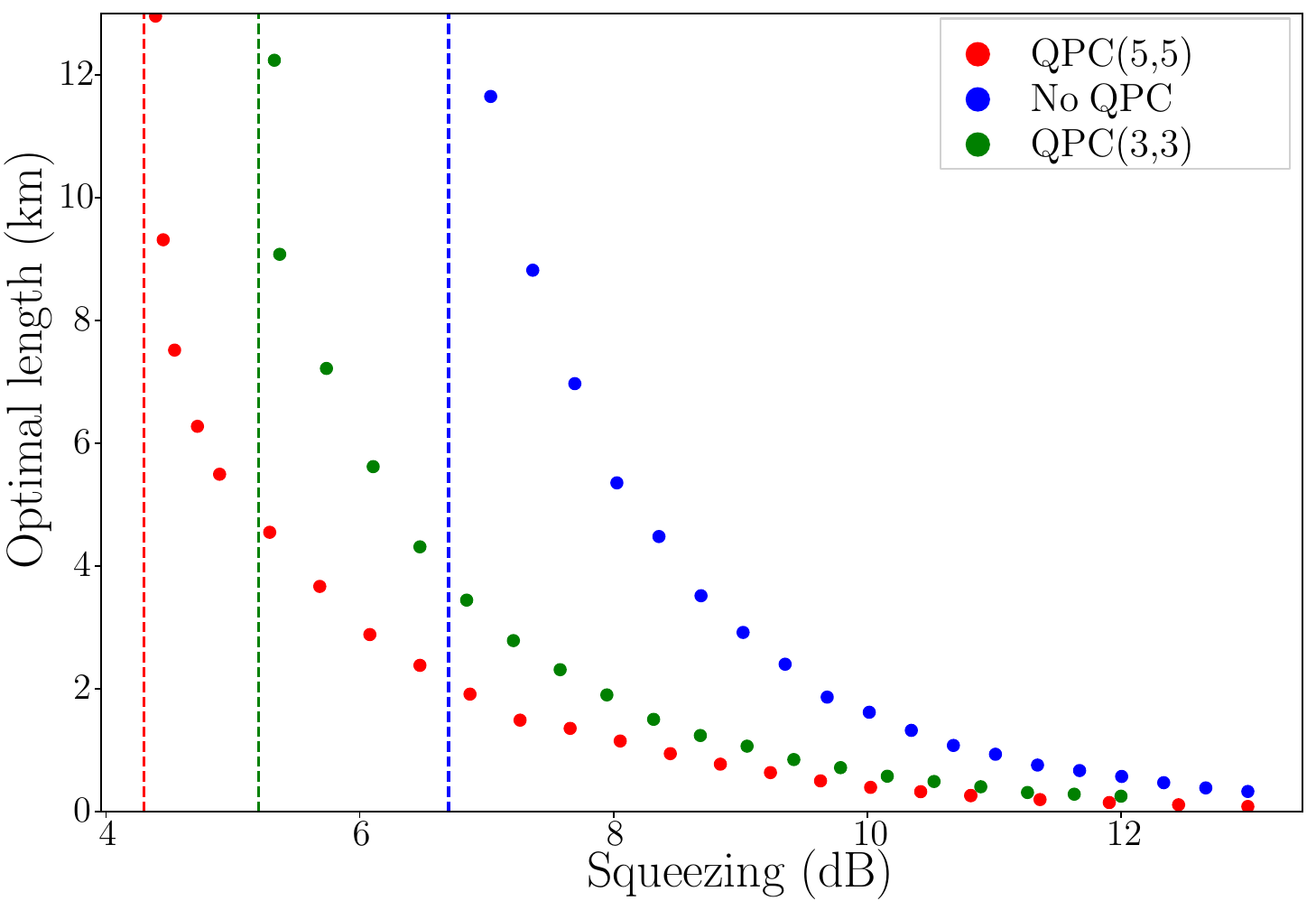}
    \caption{Optimal segment length as a function of GKP squeezing. The threshold below which repeated error correction provides no advantage shifts towards lower squeezing as the outer QPC code size increases: from $6.7\,\mathrm{dB}$ (no QPC) to $5.2\,\mathrm{dB}$ with QPC$(3,3)$ and $4.3\,\mathrm{dB}$ with QPC$(5,5)$.}
    \label{fig:optlength}
\end{figure}

The present concatenation scheme shifts this threshold to lower squeezing. As shown in Fig.~\ref{fig:optlength}, adding QPC$(3,3)$, corresponding to the Shor code,  lowers the threshold to $5.2\,\mathrm{dB}$, while QPC$(5,5)$ reduces it further to $4.3\,\mathrm{dB}$. This illustrates the core benefit of the outer code. The inner GKP layer converts loss-induced continuous-variable noise into residual rail-level Pauli errors, while the QPC layer suppresses these residual errors at the logical level. Consequently, even when individual GKP rails are below the bare-GKP threshold, their concatenation within the QPC can make repeated correction beneficial.

%concatenation: below the bare-GKP threshold, the noise introducd by each teleportation-based error correction step outweighs the benefit of correcting accumulated loss errors. The outer QPC layer acts as an additional filter on this residual noise, suppressing errors that the inner GKP layer cannot handle alone. Notably, even when individual GKP qubits are below threshold, their concatenation within the QPC can still be beneficial---which is precisely the operational principle of quantum error correction.

\subsection{Storage time and fidelity advantage}
\begin{figure}[h]
    \centering
    \includegraphics[width=1\linewidth]{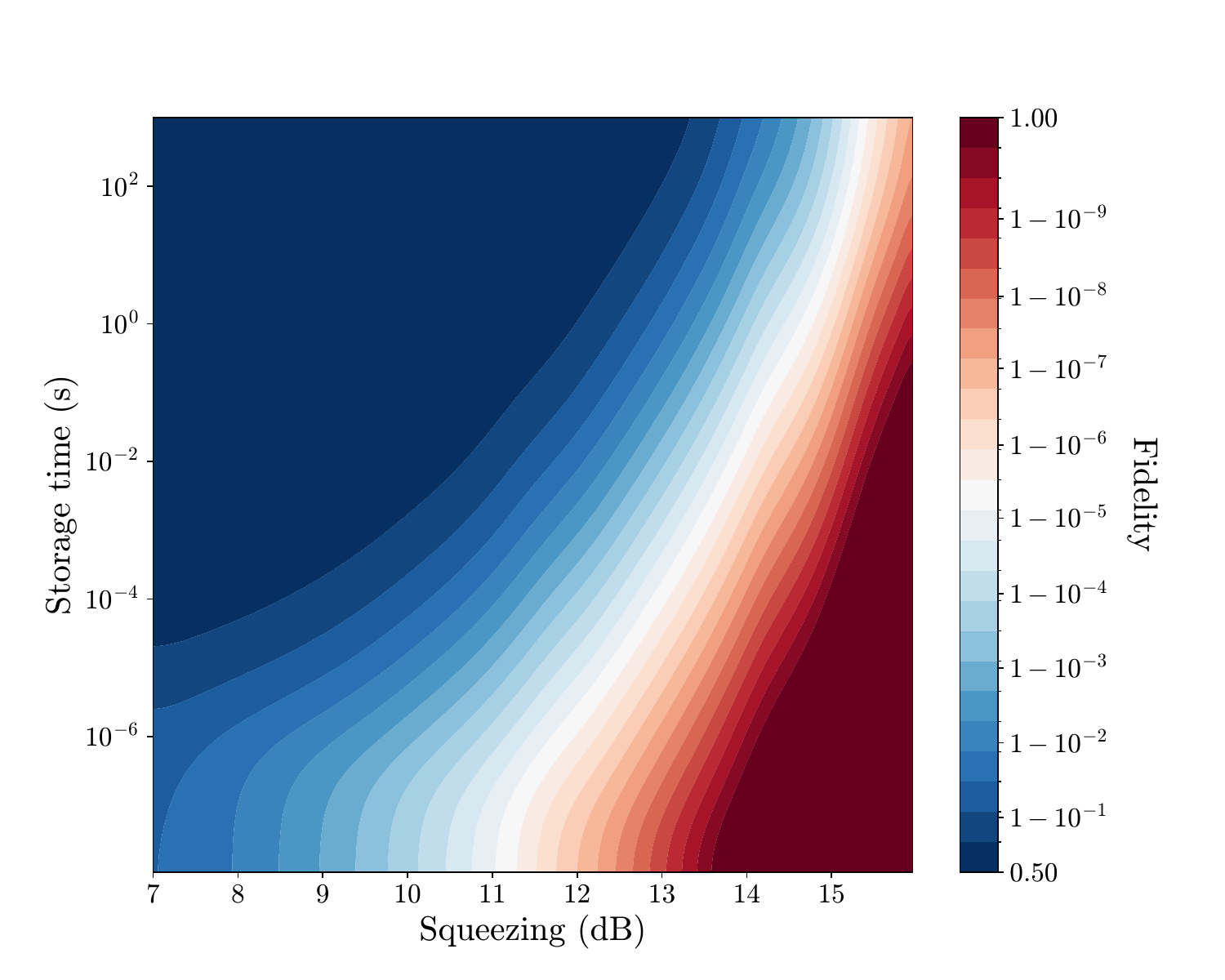}
    \caption{Average fidelity of the quantum memory with GKP--QPC$(5,5)$ as a function of GKP squeezing and storage time $t_s$. Fidelities are computed using the optimal segment lengths from Fig.~\ref{fig:optlength}.}
    \label{fig:memory5}
\end{figure}

Figure~\ref{fig:memory5} shows the average memory fidelity for GKP--QPC$(5,5)$ using the optimized segment lengths from Fig.~\ref{fig:optlength}. As expected, the fidelity increases with GKP squeezing and decreases when we demand more storage time. The broad high-fidelity region shows that, once the GKP-corrected rail noise is below the operating scale of the QPC decoder, the outer code substantially slows the logical decay.
%but the contours remain high over a broad region because the outer code delays the logical decay once the inner GKP rail noise is below the QPC operating scale.

To compare directly with the bare-GKP memory, we define the logarithmic fidelity-gain ratio
\begin{equation}
    G_{n,m}(\varepsilon,t)
    :=
    \log_{10}\!\left(
    \frac{F_{\mathrm{QPC}(n,m)}(\varepsilon,t)}
         {F_{\mathrm{GKP}}(\varepsilon,t)}
    \right),
    \label{eq:memory_gain_ratio}
\end{equation}
where positive values indicate that the concatenated memory has a higher average fidelity than the bare-GKP memory at the same squeezing and storage time. This ratio highlights the relative advantage of concatenation in the parameter regime where the absolute fidelity background remains high.

\begin{figure}[h]
    \centering
    \includegraphics[width=1\linewidth]{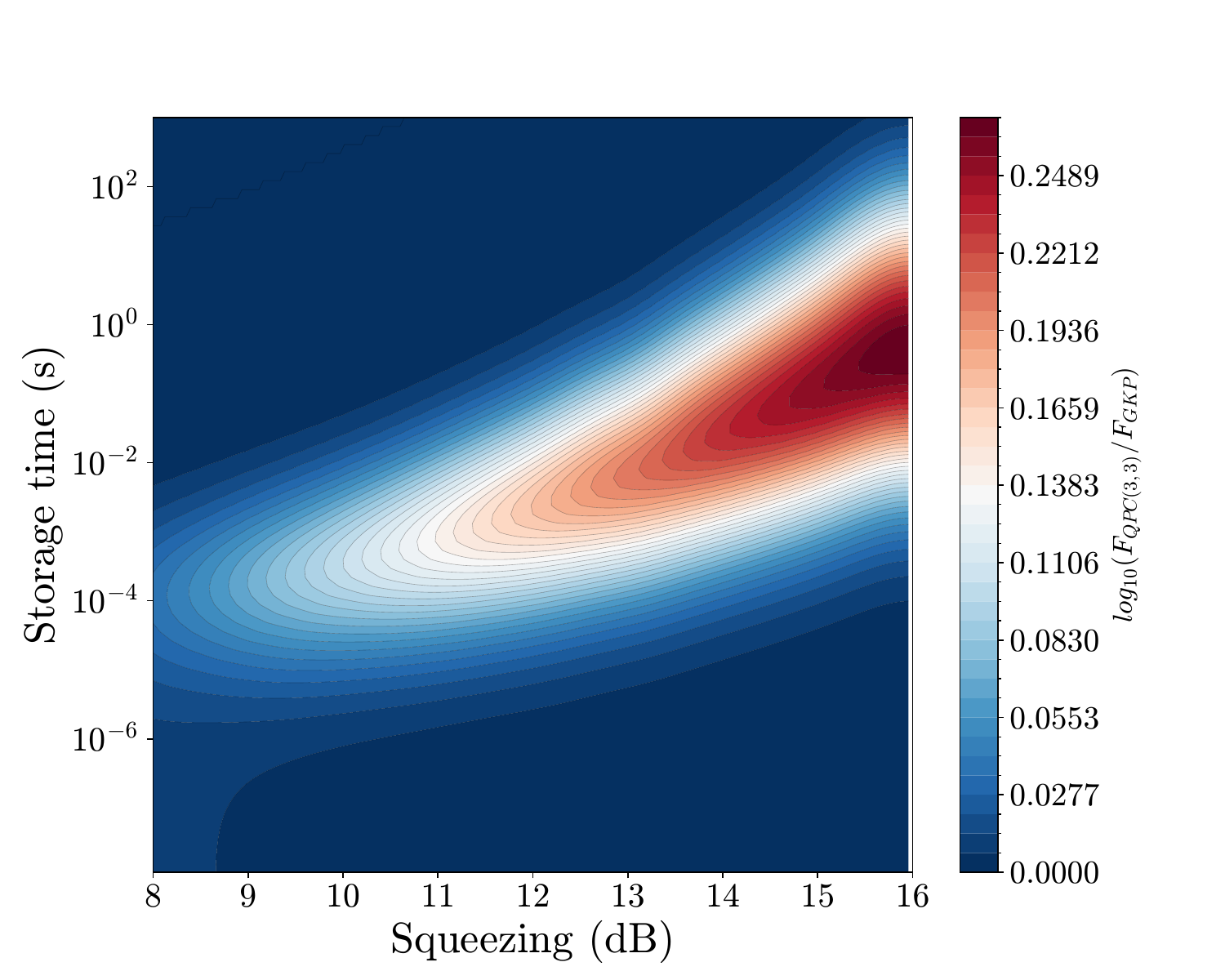}
    \caption{Logarithmic fidelity gain $G_{3,3}(\varepsilon,t)$ for GKP--QPC$(3,3)$ relative to the bare-GKP memory. The largest gain, $G_{3,3}\approx 0.25$, corresponds to a fidelity ratio of approximately $1.8$ relative to bare GKP.}
    \label{fig:memadv3}
\end{figure}

Figure~\ref{fig:memadv3} shows $G_{3,3}$ for the QPC$(3,3)$, or Shor-code concatenation. The advantageous region forms a diagonal band: as the storage time increases, higher squeezing is needed before the outer QPC layer becomes useful. The peak value is $G_{3,3}\simeq0.25$, corresponding to a fidelity ratio $F_{\mathrm{QPC}(3,3)}/F_{\mathrm{GKP}}\simeq10^{0.25}\approx1.8$.
%n approximately $75\%$ increase in the fidelity ratio relative to bare GKP. 
This behavior is physically expected. At very short storage times, the bare-GKP memory is already close to ideal, leaving little room for improvement. At very long storage times and low squeezing, the rail-level noise is too large for the small outer code to suppress.

\begin{figure}[h]
    \centering
    \includegraphics[width=1\linewidth]{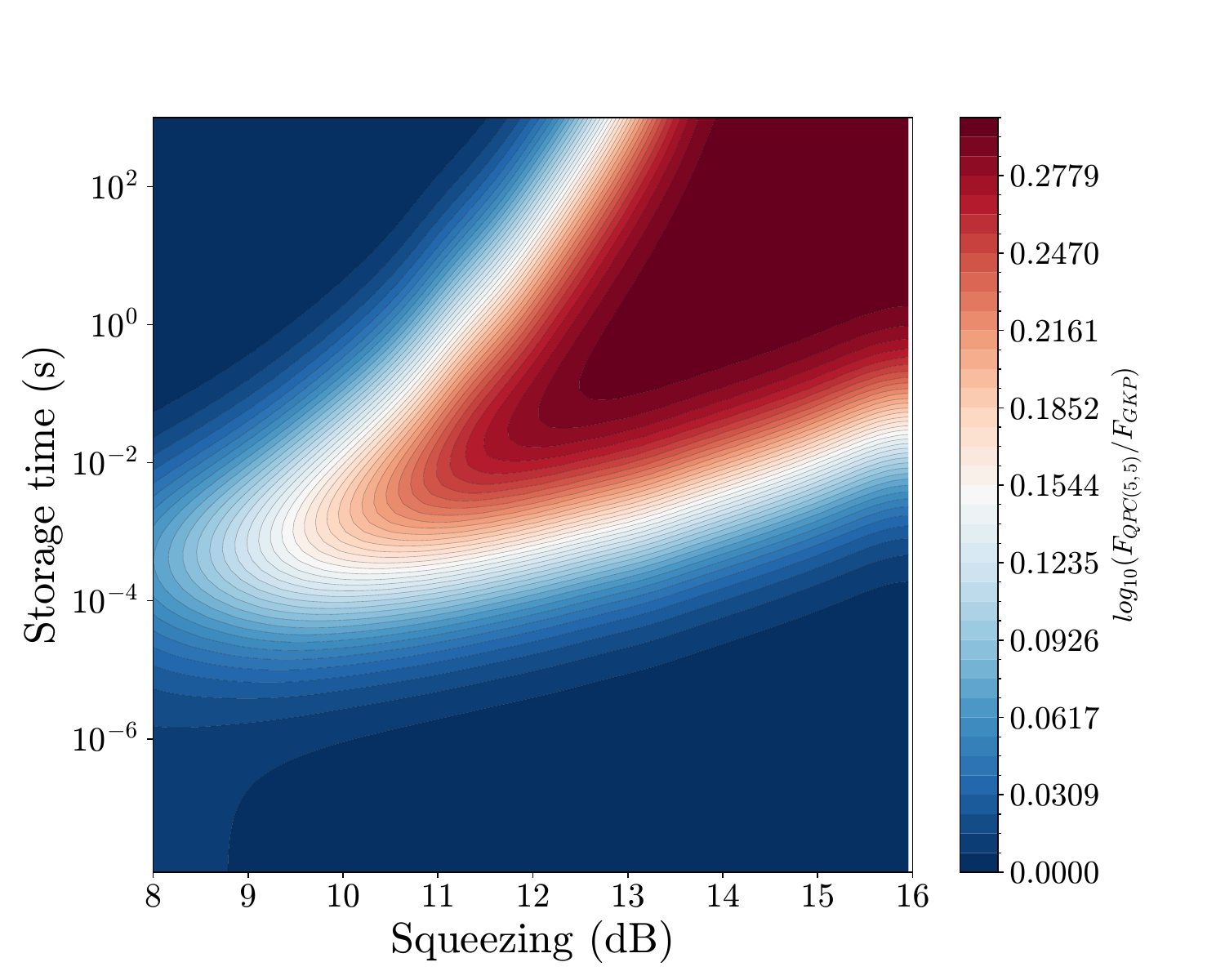}
    \caption{Logarithmic fidelity gain $G_{5,5}(\varepsilon,t)$ for GKP--QPC$(5,5)$ relative to the bare GKP memory. The peak gain of $G_{5,5}\approx 0.28$ corresponds to a fidelity ratio of approximately $1.9$. Compared with Fig.~\ref{fig:memadv3}, the advantageous region is broader and extends to longer storage times at the same squeezing.}
    \label{fig:memadv5}
\end{figure}

Figure~\ref{fig:memadv5} shows the same comparison for QPC$(5,5)$. The peak gain is only modestly larger, $G_{5,5}\simeq0.28$, corresponding to a fidelity ratio of approximately $10^{0.28}\approx1.9$. The main benefit of the larger outer code is therefore not the maximum gain itself, but the broader parameter window over which concatenation is advantageous. QPC$(5,5)$ uses more physical rails than QPC$(3,3)$, but it converts the finite-squeezing GKP layer into a more robust logical storage channel over a wider range of squeezing values and storage times.

\section{Architecture for quantum repeater}
\label{sec:architecture_rates}
We now apply the same finite-squeezing GKP--QPC correction module to a one-way quantum repeater. We consider a repeater chain of total length $L_{\mathrm{tot}}$ divided into $N_{\mathrm{hop}}$ elementary segments of length $L_{\mathrm{seg}}$,
\begin{equation}
    L_{\mathrm{tot}} = N_{\mathrm{hop}} L_{\mathrm{seg}},
\end{equation}
with segment transmissivity
\begin{equation}
    \eta(L_{\mathrm{seg}})=e^{-L_{\mathrm{seg}}/L_{\mathrm{att}}}.
\end{equation}
At each station, the incoming rail is corrected by teleportation-based GKP error correction. In the bare-GKP architecture, this gives the effective single-rail PTM $C^{\varepsilon,\eta}$. In the concatenated architecture the corresponding rail-level error statistics are instead passed to the QPC decoder, giving the logical PTM $C_L^{\varepsilon,\eta}$ derived in Sec.~\ref{sec:a2}.

\subsection{Bare-GKP repeater chain}

Let $C^{\varepsilon,\eta}$ be the accepted single-hop GKP map in the Bloch-vector representation
\begin{equation}
    v=(1,r_x,r_y,r_z)^T .
\end{equation}
After $N_{\mathrm{hop}}$ identical segments, the end-to-end map is
\begin{equation}
    (C^{\varepsilon,\eta})^{N_{\mathrm{hop}}} .
\end{equation}
To extract the bit and phase error rates entering the asymptotic BB84~\cite{Bennett_2014} secret key fraction, we probe this channel with the ideal logical $X$- and $Z$-basis Bloch vectors,
\begin{equation}
    v_X=(1,1,0,0)^T,
    \qquad
    v_Z=(1,0,0,1)^T .
\end{equation}
Defining
\begin{equation}
    \ell_X = (C^{\varepsilon,\eta})^{N_{\mathrm{hop}}}v_X,
    \qquad
    \ell_Z = (C^{\varepsilon,\eta})^{N_{\mathrm{hop}}}v_Z,
\end{equation}
the normalized output components are
\begin{equation}
    r_X^{\mathrm{out}}=\frac{(\ell_X)_1}{(\ell_X)_0},
    \qquad
    r_Z^{\mathrm{out}}=\frac{(\ell_Z)_3}{(\ell_Z)_0}.
\end{equation}
These directly determine the basis error probabilities:
\begin{equation}
    Q_X=\frac{1-r_X^{\mathrm{out}}}{2},
    \qquad
    Q_Z=\frac{1-r_Z^{\mathrm{out}}}{2}.
\end{equation}
Since the final logical Pauli frame can be chosen such that the reported error probability is the smaller of the two complementary outcomes, we use
\begin{equation}
    \widetilde Q_X=\min(Q_X,1-Q_X),
    \qquad
    \widetilde Q_Z=\min(Q_Z,1-Q_Z).
\end{equation}
The asymptotic one-way BB84 secret-key fraction (SKF) per transmitted logical qubit is then~\cite{PhysRevA.87.052315}
\begin{equation}
    r_{\mathrm{GKP}}
    =
    \max\!\left\{
    0,\,
    1-h_2(\widetilde Q_X)-h_2(\widetilde Q_Z)
    \right\},
\end{equation}
where
\begin{equation}
    h_2(q)=-q\log_2 q-(1-q)\log_2(1-q)
\end{equation}
is the binary entropy.

\subsection{Concatenated GKP--QPC repeater}

For the concatenated repeater, we apply the full GKP--QPC correction at every station. More general interleaved architectures, where the outer code is applied only after several GKP-only hops, are possible~\cite{Rozp_dek_2021} but are not optimized here. With one concatenated correction step per elementary segment, the end-to-end logical map is
\begin{equation}
    (C^{\varepsilon,\eta}_L)^{N_{\mathrm{hop}}} .
\end{equation}
The logical $X$- and $Z$-basis error rates are extracted by the same probing procedure as above, giving $\widetilde Q_X^L$ and $\widetilde Q_Z^L$. The corresponding asymptotic BB84 secret-key fraction per transmitted logical qubit is
\begin{equation}
    r_{\mathrm{QPC}}
    =
    \max\!\left\{
    0,\,
    1-h_2(\widetilde Q^L_X)-h_2(\widetilde Q^L_Z)
    \right\}.
\end{equation}

A bare-GKP repeater uses one optical mode per transmitted logical qubit, whereas QPC$(n,m)$ uses $mn$ physical rails. A comparison only at the logical-qubit level would therefore automatically favor larger outer codes. Following the resource-normalized perspective of Ref.~\cite{Rozp_dek_2021}, we report the SKF per optical mode,
\begin{equation}
    r_0
    =
    \frac{r}{n_{\mathrm{modes}}},
\end{equation}
where
\begin{equation}
    n_{\mathrm{modes}}=
    \begin{cases}
        1, & \text{bare GKP},\\[4pt]
        mn, & \text{GKP--QPC}.
    \end{cases}
\end{equation}
This normalization is physically important: it isolates the genuine coding advantage from the trivial gain that would otherwise arise from simply increasing redundancy. In particular, it allows one to decide whether a larger outer code improves long-distance performance \emph{per occupied optical mode}, rather than merely per encoded logical qubit.

The overall repeater model can be summarized as follows. The inner GKP layer converts each elementary pure-loss segment into an effective qubit channel. In the bare-GKP architecture, these single-rail maps are concatenated along the chain, and the BB84 error rates are read off from the final $X$- and $Z$-basis error rates. In the GKP--QPC architecture, the same rail-wise GKP maps feed an outer QPC decoder, which converts correct/incorrect sector statistics into logical error rates $E_X$ and $E_Z$ through the hard-decoder combinatorics. These logical error rates define the effective logical PTM $C_L^{\varepsilon,\eta}$, which is then propagated along the repeater chain. Both architectures are thus evaluated within the same resource-normalized framework. The BB84 secret-key fraction is finally extracted from the end-to-end logical matrix and normalized by the number of optical modes used. Both the bare-GKP and the concatenated GKP--QPC repeaters are thereby evaluated within a common, resource-normalized framework.
\subsubsection{Achievable rates: A comparative view}
The overall idea of using GKP codes concatenated with outer codes is not new and certain rate per mode calculations were presented as discussed in \cite{Rozp_dek_2021}. Here we analyze how much variation our model produces within a similar setting of allowing for minimum repeater placement of $250$m as done in \cite{Rozp_dek_2021}. We use the same reporting metric: the largest total distance for which the secret-key fraction per transmitted optical mode remains at least $10^{-2}$. For an elementary link of length $L_{\mathrm{seg}}$, the coupling efficiency is included once per hop through
\begin{equation}
    \begin{aligned}
        \eta_{\mathrm{hop}}(L_{\mathrm{seg}})
        &=\eta_{\mathrm{c}}\,\eta_L(L_{\mathrm{seg}})\\
    \end{aligned}
    \label{eq:coupling_aware_hop_transmissivity}
\end{equation}
where $\eta_{\mathrm{c}}$ is the coupling efficiency of single segment. The tables for the maximum reach achieved is provided in \ref{tab:pure_loss_gkp_reach} and \ref{tab:pure_loss_qpc33_reach} which are optimized over a segment length $L_{seg}\in[0.25,1.5]$km chosen based on \cite{Rozp_dek_2021}. In both tables, $<100$ means that the threshold is crossed below the lower reporting scale of $100\,\mathrm{km}$, while $>10\,000$ means that the rate
remains above threshold at the $10^4\,\mathrm{km}$ reporting ceiling. The tables shows the expected qualitative trend where higher squeezing, higher $\eta_c$ facilitates larger reach and code concatenation improves it. But the values obtained are in general comparable or even at times improved over the figure 3 in \cite{Rozp_dek_2021} with up-to two order of magnitude difference. Given that the QPC$(3,3)$ is larger code than Steane code this is expected to some degree however we still consider per optical mode performance to make comparisons more leveled.

\begin{table}[t]
    \caption{Maximum reach $L_{\max}$ in kilometres for the bare-GKP repeater
    in the present unamplified pure-loss model.  Columns give the per-hop coupling efficiency $\eta_{\mathrm{c}}$.}
    \label{tab:pure_loss_gkp_reach}
    \centering
    \begin{ruledtabular}
        \begin{tabular}{ccccc}
            GKP squeezing & \multicolumn{4}{c}{$\eta_{\mathrm{c}}$} \\
            (dB) & $0.93$ & $0.95$ & $0.97$ & $0.99$ \\
            \hline
            $17.9$ & $<100$ & $<100$ & $1320$ & $>10\,000$ \\
            $16.2$ & $<100$ & $<100$ & $448$  & $>10\,000$ \\
            $14.7$ & $<100$ & $<100$ & $131$  & $951$
        \end{tabular}
    \end{ruledtabular}
\end{table}
\begin{table}[t]
    \caption{Maximum reach $L_{\max}$ in kilometres for the GKP--QPC$(3,3)$
    repeater in the present unamplified pure-loss model.  The BB84 secret-key
    fraction is divided by the nine transmitted GKP modes before applying the
    threshold $r_0\geq10^{-2}$.}
    \label{tab:pure_loss_qpc33_reach}
    \centering
    \begin{ruledtabular}
        \begin{tabular}{ccccc}
            GKP squeezing & \multicolumn{4}{c}{$\eta_{\mathrm{c}}$} \\
            (dB) & $0.93$ & $0.95$ & $0.97$ & $0.99$ \\
            \hline
            $17.9$ & $283$ & $>10\,000$ & $>10\,000$ & $>10\,000$ \\
            $16.2$ & $321$ & $5693$      & $>10\,000$ & $>10\,000$ \\
            $14.7$ & $192$ & $1674$      & $>10\,000$ & $>10\,000$
        \end{tabular}
    \end{ruledtabular}
\end{table}

\subsection{Results}

\subsubsection{Advantage of forgoing pre-amplification}

In the main architecture considered here, we do not apply phase-insensitive pre-amplification to compensate for channel loss before GKP correction. Such pre-amplification maps pure loss to an effective random-displacement channel, but it also changes the balance of finite-squeezing errors. For a single round of GKP error correction, Ref.~\cite{PhysRevA.108.052413} showed that avoiding pre-amplification gives higher average fidelity at moderate squeezing, whereas pre-amplification can become favorable at very high squeezing. Since experimentally available GKP squeezing is likely to be limited, the moderate-squeezing regime is particularly relevant. Here we examine the same tradeoff in the repeater setting.

%Using pure loss as the transmission noise model means that we do not amplify the state at each repeater station before passing it to the next node. It has been shown~\cite{PhysRevA.108.052413} that, for a single round of GKP error correction, operating without pre-amplification yields higher average fidelity at moderate squeezing values than the amplified case, which maps the channel to a random-displacement channel. This is particularly significant since higher GKP squeezing is an experimental bottleneck. Here we extend this result to the repeater setting.

\begin{figure}[h]
    \centering
    \includegraphics[width=1\linewidth]{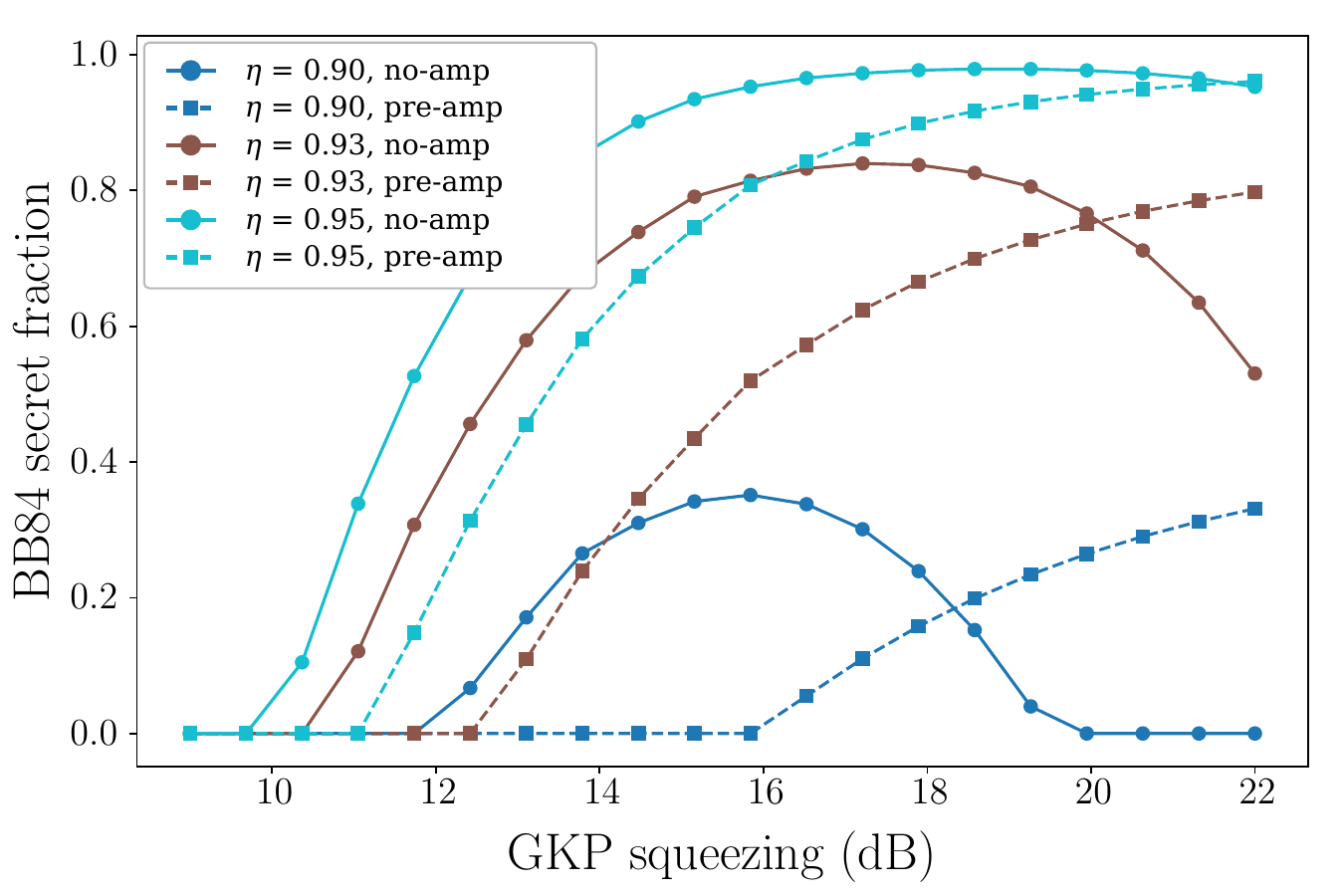}
    \caption{Secret-key fraction as a function of GKP squeezing with and without pre-amplification, evaluated after 10 repeater stations for several per-segment transmissivities $\eta$. These are for bare GKP repeater stations without any outer code. Without pre-amplification, the SKF is higher at moderate squeezing but decreases at high squeezing due to peak-shifting errors, leading to an optimal GKP squeezing for each $\eta$.}
    \label{fig:preamp}
    \end{figure}

\begin{figure}[h]
    \centering
  \includegraphics[width=1\linewidth]{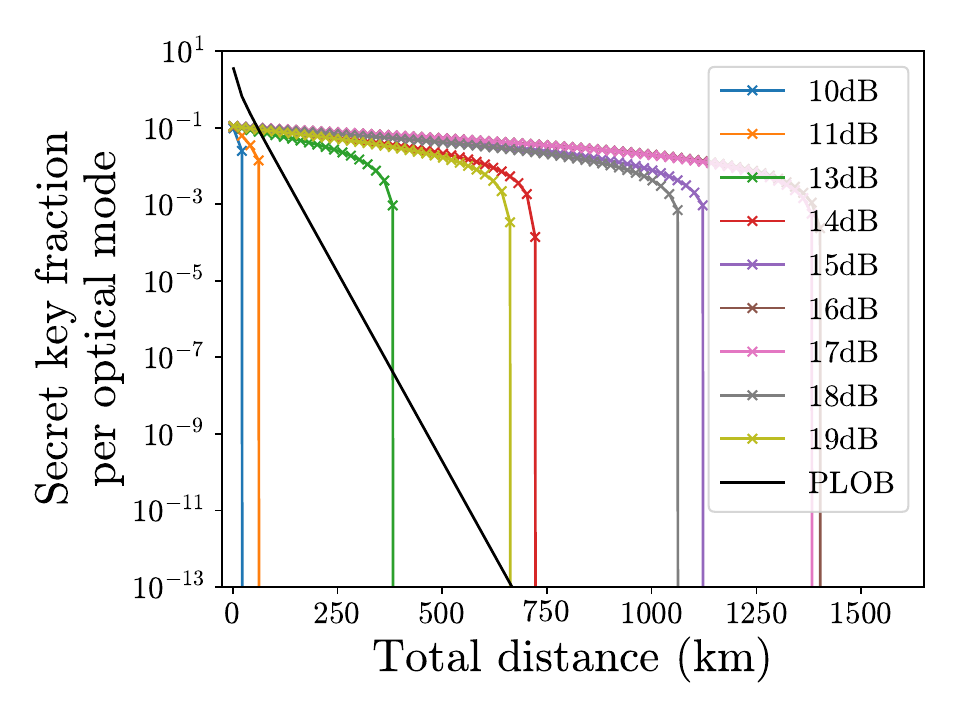}
    \caption{Secret-key fraction for the BB84 protocol using GKP--QPC$(3,3)$ as a function of total distance, for several GKP squeezing values. The repeater spacing is $2\,\mathrm{km}$. Even at moderate squeezing of $14\,\mathrm{dB}$ the SKF significantly exceeds the PLOB bound (solid black line). The $19\,\mathrm{dB}$ curve lying below the $14\,\mathrm{dB}$ curve is a consequence of peak-shifting errors in the absence of pre-amplification; the per-mode SKF is optimized near $16\,\mathrm{dB}$.}
    \label{fig:qpc33preamp}
    \end{figure}

Fig.~\ref{fig:preamp} shows the SKF as a function of GKP squeezing for several different per-segment transmissivities $\eta$ after $N_{\mathrm{hop}}=10$ elementary segments. Without pre-amplification, the SKF is larger at moderate squeezing, but it reaches an optimum and then decreases. This non-monotonic behavior reflects two competing effects: At low squeezing, increased Gaussian peak variance due to added vacuum noise is the dominant error mechanism. At high squeezing, inward displacement of the peaks far from the origin dominates. Pre-amplification eliminates the inward displacement by restoring the peak positions, thereby improving the high-squeezing behavior at the cost of the moderate-squeezing advantage.
Adding the outer QPC code allows the repeater to exceed the PLOB bound~\cite{Pirandola_2017} when the elementary segments are sufficiently short. This does not violate the PLOB bound, which applies to direct repeaterless transmission over the full distance; rather, it demonstrates the expected advantage of an active repeater chain. Figure~\ref{fig:qpc33preamp} shows the SKF for GKP--QPC$(3,3)$ with repeater spacing $L_{\mathrm{seg}}=2\,\mathrm{km}$. Already at $14\,\mathrm{dB}$ squeezing, the SKF lies well above the repeaterless PLOB benchmark over the plotted distance range.  The no-amplification optimum is again visible: the $19\,\mathrm{dB}$ curve lies below the $14\,\mathrm{dB}$ curve because peak-shifting errors dominate at high squeezing, while the best per-mode SKF occurs near $16\,\mathrm{dB}$.

%Here too, the no-amplification advantage is visible: the $19\,\mathrm{dB}$ curve falls below the $14\,\mathrm{dB}$ curve due to peak-shifting errors, with the per-mode SKF optimized near $16\,\mathrm{dB}$. This confirms the existence of an optimal squeezing that mirrors the optimal segment-length behavior found in the quantum memory case.

\subsubsection{Outer-code reach and segment-length threshold}

\begin{figure}[h]
    \centering
    \includegraphics[width=1\linewidth]{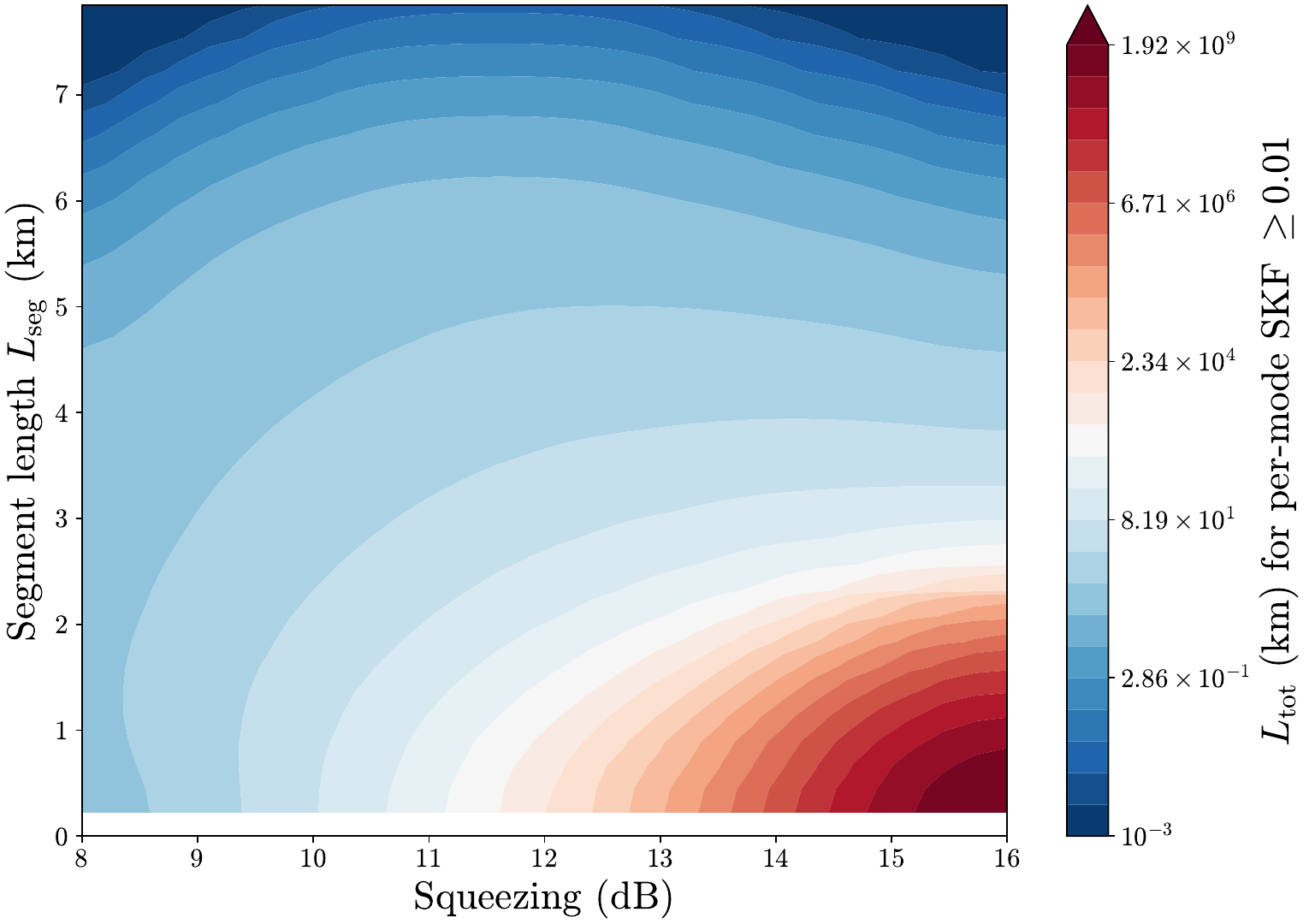}
    \caption{Maximum total communication distance for which the per-mode SKF exceeds $0.01$, as a function of repeater spacing $L_\mathrm{seg}$ and GKP squeezing, using GKP--QPC$(3,3)$. At a repeater spacing of $1\,\mathrm{km}$ and squeezing of $14\,\mathrm{dB}$, the Shor code enables reaching distances of order $10^4\,\mathrm{km}$.}
    \label{fig:optrepseg}
\end{figure}

Figure~\ref{fig:optrepseg} addresses a network-design question: for a fixed repeater spacing $L_\mathrm{seg}$ and GKP squeezing $\varepsilon$, how far can the chain be extended before the per mode SKF falls below $0.01$? This threshold is not a fundamental capacity boundary, but an operational benchmark that allows different designs to be compared at a fixed useful key fraction. The contour plot shows three main features. First, at fixed $L_\mathrm{seg}$, the reach improves rapidly as the GKP squeezing increases from the low-squeezing regime because the teleportation-based correction leaves a cleaner rail-level qubit channel for the QPC decoder. Second, the improvement eventually saturates and may reverse: in the no-amplification architecture, highly squeezed peaks become more sensitive to loss-induced inward displacement discussed above, so increasing the squeezing is not always beneficial. Third, the reach is strongly controlled by the spacing, since the single-hop transmissivity decreases exponentially with $L_\mathrm{seg}$. For example, at  $L_\mathrm{seg}=1\,\mathrm{km}$ and $14\,\mathrm{dB}$ squeezing, QPC$(3,3)$ maintains a per-mode SKF above $0.01$ out to distances of order larger than $10^4\,\mathrm{km}$.

The absolute reach in Fig.~\ref{fig:optrepseg} does not by itself show whether the QPC layer is responsible for the improvement, because the bare-GKP repeater also benefits from short segments and higher squeezing. We therefore compare the concatenated repeater directly with the bare-GKP repeater through the logarithmic reach gain
\begin{equation}
    \Delta(\varepsilon,L_{\mathrm{seg}})
    :=
    \log_{10}\!\left(
    \frac{L_{\mathrm{QPC}}(\varepsilon,L_{\mathrm{seg}})}
         {L_{\mathrm{GKP}}(\varepsilon,L_{\mathrm{seg}})}
    \right),
    \label{eq:delta_reach_ratio}
\end{equation}
where $L_{\mathrm{QPC}}$ and $L_{\mathrm{GKP}}$ are the maximum distances for which the corresponding per-mode SKF remains above $0.01$. Positive values of $\Delta$ mean that the QPC layer extends the useful distance; negative values mean that, after mode normalization, the outer code has reduced the useful distance.

\begin{figure}[h]
    \centering
    \includegraphics[width=1\linewidth]{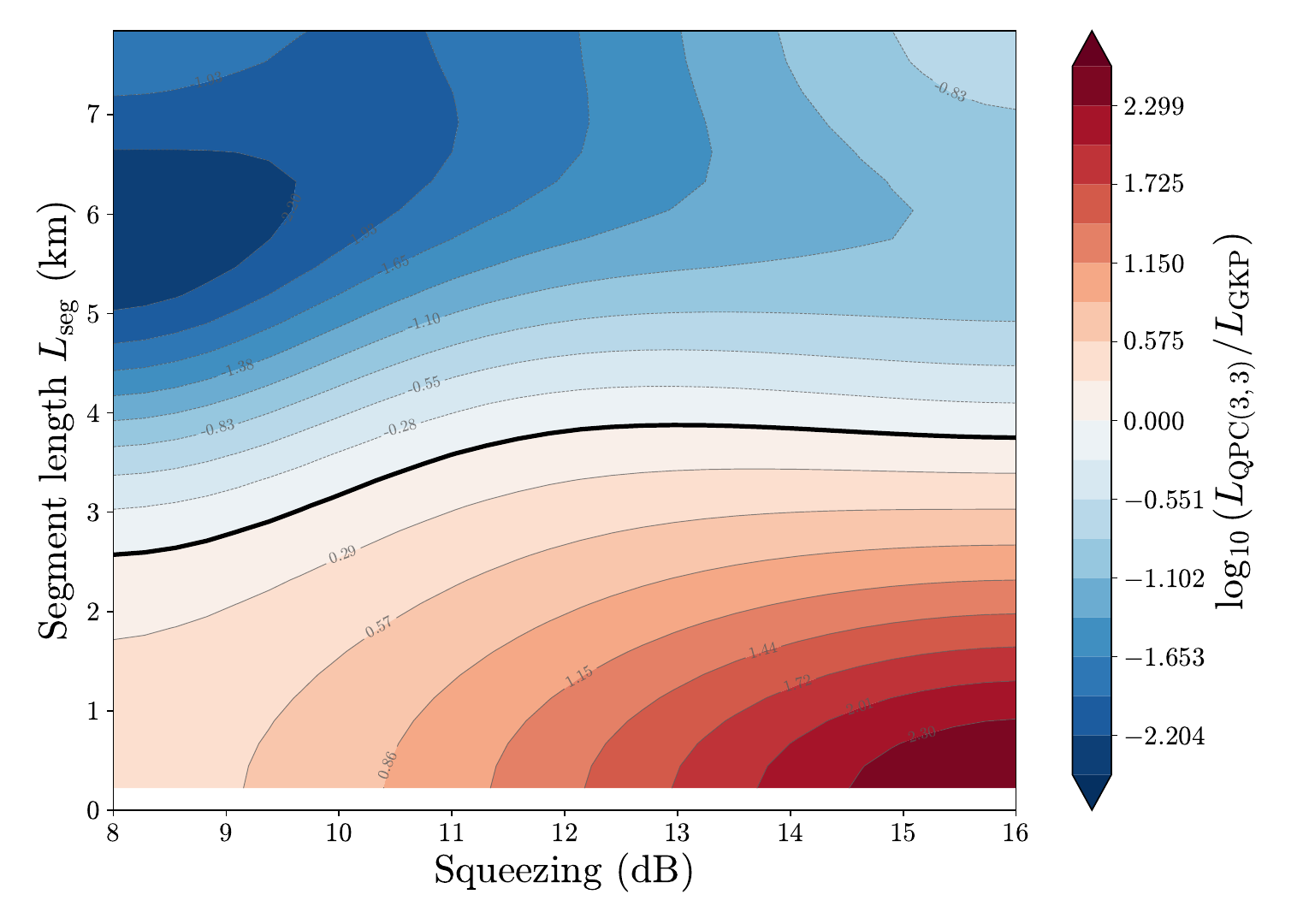}
    \caption{Logarithmic gain $\Delta(\varepsilon,L_\mathrm{seg})$ in maximum reach: GKP--QPC$(3,3)$ versus bare GKP. Positive regions indicate a reach advantage from concatenation; negative regions indicate that the outer code is detrimental. The bold black contour at $\Delta=0$ marks the threshold segment length beyond which using QPC$(3,3)$ becomes harmful.}
    \label{fig:repadv}
\end{figure}

Figure~\ref{fig:repadv} shows that the advantage of concatenation is sharply regime-dependent. The positive region occurs at short segment lengths, where each elementary hop has high transmissivity and the inner GKP correction already produces a reasonably clean effective qubit channel. In this regime, the bare GKP repeater is limited mainly by the accumulation of many small residual logical errors over a long chain. The QPC layer is well matched to this situation: it suppresses these rail-level errors before they accumulate, and the resulting reduction compounds over many repeater stations. This is why the strongest positive gain appears in the short-spacing, high-squeezing part of the plot.

For larger segment lengths the situation reverses. Each rail then enters the GKP correction step after a significantly more lossy hop, so the physical error marginals supplied to the QPC decoder are already too poor. Once the rail-level channel is outside the useful operating regime of the outer code, majority and parity processing cannot compensate for the degraded physical transmissions. The concatenated architecture then pays the resource-normalization penalty associated with using $mn$ optical modes, while gaining too little logical suppression in return. This produces the negative region of Fig.~\ref{fig:repadv}, where the bare GKP repeater has the larger resource-normalized reach.

The zero contour $\Delta=0$ is therefore a practical segment-length threshold for the chosen outer code. Below this contour, QPC$(3,3)$ is a useful concatenation layer; above it, the elementary hop is too noisy for this small outer code to help. Increasing the GKP squeezing shifts the contour upward because a cleaner inner-GKP channel keeps the QPC decoder below its useful  operating regime for larger segment lengths. The shift eventually flattens, however, because transmission loss and peak shifting become the limiting mechanisms. Thus Fig.~\ref{fig:repadv} gives a design rule: QPC concatenation should be used in the clean-hop regime where the inner GKP layer has already reduced the noise to a correctable level. For longer hops, one should either reduce the repeater spacing, use a different outer-code strategy, or revert to the bare GKP-architecture.

\subsubsection{Optimization over the outer code}

We finally optimize the QPC parameters at fixed total distance and repeater spacing. For $L_{\mathrm{tot}}=10^4\,\mathrm{km}$ and $L_{\mathrm{seg}}=2\,\mathrm{km}$, we scan over $(n,m)$ and maximize the per-mode SKF,
\begin{equation}
    (n^\star,m^\star)(\varepsilon)
    =
    \arg\max_{n,m}\, r_{\mathrm{QPC}}(\varepsilon;n,m).
\end{equation}
The normalization by $mn$ is crucial: the optimizer is not allowed to improve the objective merely by using a larger code.

\begin{figure}[h]
    \centering
    \includegraphics[width=1\linewidth]{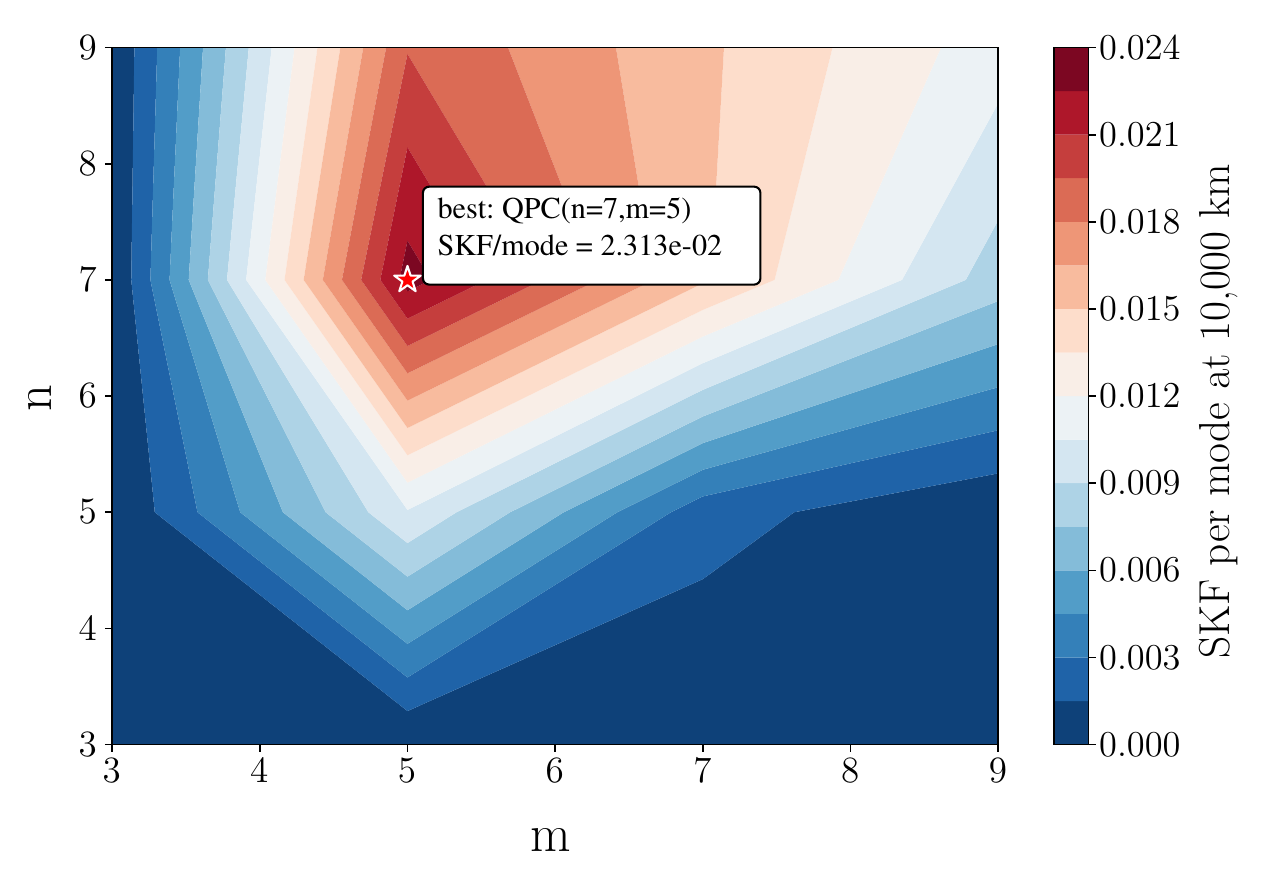}
    \caption{Secret-key fraction per mode as a function of QPC parameters for a fixed repeater spacing of $2\,\mathrm{km}$, GKP squeezing of $16\,\mathrm{dB}$, and total distance of $10^4\,\mathrm{km}$. In the manuscript convention QPC$(n,m)$ has $n$ blocks of $m$ physical qubits; the plotted axes store the same information as $(m,n)$. The optimum corresponds to QPC$(7,5)$, yielding a per-mode SKF of $\approx 0.023$ and a total SKF of $\approx 0.8$ across $35$ optical modes.}
    \label{fig:mn16db}
\end{figure}

Figure~\ref{fig:mn16db} shows the rate landscape at $16\,\mathrm{dB}$. The optimum is not at the largest tested code but at an interior point, QPC$(7,5)$, with SKF $\simeq 2.313\times 10^{-2}$. This interior optimum reflects the resource tradeoff. If the code is too small, residual rail errors are not suppressed enough over $10^4\,\mathrm{km}$. If the code is too large, the logical improvement is outweighed by the $mn$-mode overhead. Nearby codes achieve comparable rates, indicating that the optimum is robust rather than a fine-tuned point.

\begin{figure}[h]
    \centering
    \includegraphics[width=1\linewidth]{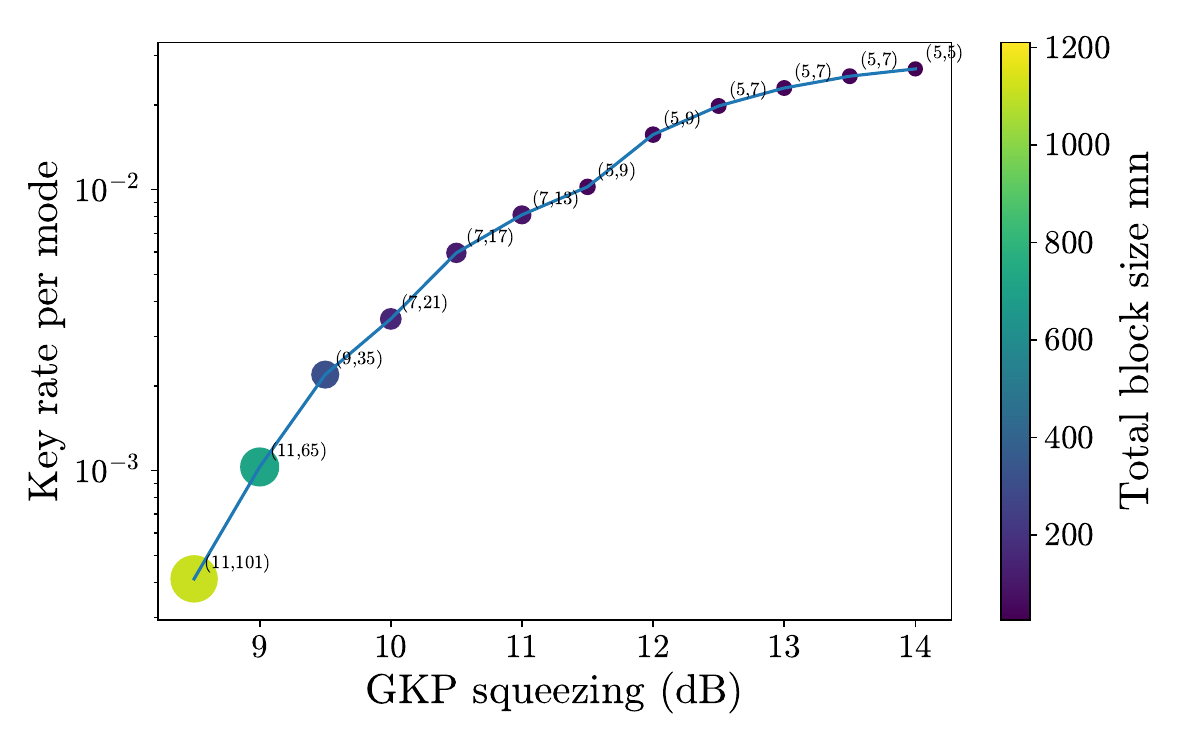}
    \caption{Optimal QPC code $(n^\star,m^\star)$ and corresponding per-mode SKF (left axis, circles) versus GKP squeezing, for a total distance of $10^3\,\mathrm{km}$ and repeater spacing of $2\,\mathrm{km}$. The right axis (color) shows the total block size $nm$. At $10\,\mathrm{dB}$ the optimal code is QPC$(21,7)$, yielding a per-mode rate of $\approx 10^{-3}$; the corresponding total SKF over all $147$ modes is of order $0.5$.}
    \label{fig:mn1000}
\end{figure}

To show how the optimal code changes with squeezing, we also perform a separate scan at the shorter target distance $L_{\mathrm{tot}}=10^3\,\mathrm{km}$, again with $L_{\mathrm{seg}}=2\,\mathrm{km}$. The result is shown in Fig.~\ref{fig:mn1000}. The optimized per-mode rate increases with squeezing, while the optimal block size decreases. At low squeezing, the inner GKP layer leaves substantial residual rail noise, so the optimizer selects large QPCs despite their mode cost. At higher squeezing, the physical GKP channel is cleaner and heavy redundancy is no longer resource-efficient; the optimum moves to smaller low-overhead codes.

\section{Conclusion}

We have proposed and benchmarked concatenated GKP--QPC architectures for photonic quantum memories and one-way quantum repeaters under finite GKP squeezing and pure-loss transmission. In this architecture, teleportation-based GKP error correction converts each lossy optical rail into an effective qubit-level channel, while the outer QPC layer suppresses the residual rail-level errors. This provides a common framework for analyzing memory fidelity, repeater secret-key fraction, optical-mode overhead, GKP squeezing, and repeater spacing within the same finite-squeezing model.

At zero propagation loss, where finite GKP ancilla squeezing is the only noise source, we find a concatenated-code threshold of $5.06\,\mathrm{dB}$. This value is higher than the ideal-ancilla code-capacity threshold of Ref.~\cite{Fukui_2023}, as expected from the additional noise introduced by finite-squeezing teleportation-based recovery. For quantum memories, the QPC layer lowers the onset of useful repeated error correction from $6.7\,\mathrm{dB}$ for bare GKP correction to $5.2\,\mathrm{dB}$ with QPC$(3,3)$ and $4.3\,\mathrm{dB}$ with QPC$(5,5)$. In the intermediate-noise regime, where the inner GKP layer is sufficiently reliable for the outer code to operate but the bare-GKP memory still decays appreciably, concatenation improves the average-fidelity ratio by roughly $75$--$90\%$.

For repeaters, we find that avoiding pre-amplification is beneficial at moderate squeezing because the elementary channel remains pure-loss rather than being converted into an effective random-displacement channel. This advantage is not monotonic: at high squeezing, loss-induced inward peak shifting produces an optimal squeezing beyond which the secret-key fraction decreases. With the outer QPC included, the repeater can exceed the repeaterless PLOB bound~\cite{Pirandola_2017} by orders of magnitude in the short-spacing regime. In particular, QPC$(3,3)$ achieves per-mode secret-key fractions above $10^{-2}$ over distances of order $10^4\,\mathrm{km}$ for short repeater spacings and moderate GKP squeezing.

The QPC layer is nevertheless not universally advantageous. Comparing directly with the bare GKP repeater reveals a segment-length threshold: for short elementary links, the outer code increases the maximum useful distance, while for long links it becomes detrimental after normalization by the number of optical modes. Physically, long segments make each rail too noisy before the QPC decoder acts, so the logical error suppression no longer compensates for the $mn$-mode overhead. Optimizing over QPC parameters confirms this resource tradeoff. For a $10^4\,\mathrm{km}$ link with $2\,\mathrm{km}$ repeater spacing and $16\,\mathrm{dB}$ squeezing, the best per-mode rate occurs at the interior optimum QPC$(7,5)$, giving an SKF per mode of approximately $2.3\times10^{-2}$.

Overall, these results show that finite-squeezing GKP--QPC concatenation can substantially improve photonic quantum memories and one-way repeaters, but only within a sharply regime-dependent operating window. The outer QPC layer is most valuable when the elementary GKP-corrected rails are already sufficiently reliable, allowing QPC decoding to convert moderate physical reliability into a large logical advantage. Outside this regime, the optical-mode overhead outweighs the logical error suppression. 

Several natural extensions remain. These include erasure-aware decoding, soft-decision use of the analog GKP syndrome information\cite{Fukui_2017,Fukui_2018,Noh_2022}, finite-efficiency homodyne detection, switching loss, imperfect Gaussian operations, realistic GKP-state preparation, and hybrid second-/third-generation repeater schemes. In such hybrid architectures, probabilistic elementary links would require quantum memories to store successfully corrected states until entanglement swapping can be performed, thereby combining the memory and repeater functions studied here~\cite{hler2024quantumrepeatersbasedstationary}. 
%Such schemes would require quantum memories to store a state until a successful entanglement swapping. This requirement would combine together the two methods proposed to build the repeater architecture.
\section{Acknowledgment}
We gratefully acknowledge support from the Danish National Research Foundation, Center for Macroscopic Quantum States (bigQ, DNRF0142), EU project CLUSTEC (grant agreement no. 101080173), EU ERC project ClusterQ (grant agreement no. 101055224, ERC-2021-ADG), European Union’s Horizon Europe research and innovation programme under the project “Quantum Security Networks Partnership” (QSNP, grant agreement no. 101114043), and Innovation Fund Denmark (CyberQ, grant agreement no. 3200-00035B, AccessQKD, grant agreement
no. 43560-00004B, PhotoQ project, grant no. 1063-00046A).

\appendix

\section{Odd-size hard-decision GKP--QPC decoder}
\label{app:qpc_hard_decoder}

This appendix states the classical decoder used to convert the single-rail GKP
channel into the logical QPC channel.  We use the convention
QPC$(n,m)$, consisting of $n$ blocks with $m$ GKP-encoded optical rails in
each block.  The numerical routines pass the block width $m$ before the
number of blocks $n$, but the code is always named QPC$(n,m)$ here.  In all
results reported in this work, both $n$ and $m$ are odd positive integers.
All majority votes below are therefore decisive in the deterministic
no-erasure setting. 
\subsection{Rail-level hard decisions}
\label{app:qpc_rail_input}

For each physical rail, teleportation-based GKP recovery uses the homodyne
outcome $(q_0,p_0)$ to select a hard Pauli-frame update.  After that update is
applied or tracked, the remaining one-rail channel is
$C^{\varepsilon,\eta}$.  Let $\alpha\in\{I,X,Y,Z\}$ denote a residual Pauli
event of this corrected rail channel. Then the coarse grained events are given by $(s_Z(\alpha),s_X(\alpha))$
\begin{align}
    s_X(\alpha)
    &=
    \begin{cases}
        +1, & \alpha\in\{I,X\},\\
        -1, & \alpha\in\{Z,Y\},
    \end{cases}\\
    s_Z(\alpha)
    &=
    \begin{cases}
        +1, & \alpha\in\{I,Z\},\\
        -1, & \alpha\in\{X,Y\}.
    \end{cases}
    \label{eq:app_qpc_rail_signs}
\end{align}
Thus a $Z$ or $Y$ frame error reverses an $X$-basis outcome, whereas an $X$
or $Y$ frame error reverses a $Z$-basis outcome.

The analytic implementation obtains the probabilities of these signs from
the one-rail PTM $C^{\varepsilon,\eta}$.  With
\begin{align}
    v_X&=(1,1,0,0)^T,
    &
    v_Z&=(1,0,0,1)^T,
\end{align}
we define
\begin{align}
    r_X
    &=\frac{\bigl(C^{\varepsilon,\eta}v_X\bigr)_1}
            {\bigl(C^{\varepsilon,\eta}v_X\bigr)_0},
    &
    r_Z
    &=\frac{\bigl(C^{\varepsilon,\eta}v_Z\bigr)_3}
            {\bigl(C^{\varepsilon,\eta}v_Z\bigr)_0}.
    \label{eq:app_qpc_rail_bloch}
\end{align}
The correct- and incorrect-sign probabilities supplied to the two decoder
branches are then
\begin{align}
    c_B&=\frac{1+r_B}{2},
    &
    i_B&=\frac{1-r_B}{2},
    \qquad B\in\{X,Z\}.
    \label{eq:app_qpc_rail_probabilities}
\end{align}
For a Pauli-diagonal rail channel these are equivalently
\begin{align}
    c_X&=p_I+p_X, & i_X&=p_Z+p_Y,\\
    c_Z&=p_I+p_Z, & i_Z&=p_X+p_Y.
    \label{eq:app_qpc_pauli_marginals}
\end{align}
as provided in \ref{corrincorr}. The reported simulations use deterministic hard decisions, so
$c_B+i_B=1$ and no rail is discarded.

\subsection{Single-shot parity and majority rule}
\label{app:qpc_single_shot}

Let $s_{B,ij}$ be the sign in basis $B$ for rail $j$ of block $i$.  The two
logical-basis decoders use the same operations in opposite order.  In the
$X$ branch, the $m$ signs in each block are first multiplied,
\begin{equation}
    u_i^{(X)}=\prod_{j=1}^{m}s_{X,ij},
    \label{eq:app_qpc_x_block_sign}
\end{equation}
and the logical sign is the majority of the $n$ block parities,
\begin{equation}
    S_X=\operatorname{sgn}\!\left(\sum_{i=1}^{n}u_i^{(X)}\right).
    \label{eq:app_qpc_x_shot}
\end{equation}
In the $Z$ branch, one first takes a majority within each block,
\begin{equation}
    u_i^{(Z)}
    =\operatorname{sgn}\!\left(\sum_{j=1}^{m}s_{Z,ij}\right),
    \label{eq:app_qpc_z_block_sign}
\end{equation}
and then multiplies the $n$ block decisions,
\begin{equation}
    S_Z=\prod_{i=1}^{n}u_i^{(Z)}.
    \label{eq:app_qpc_z_shot}
\end{equation}
Because $n$ and $m$ are odd, none of the sums in
Eqs.~\eqref{eq:app_qpc_x_shot} and \eqref{eq:app_qpc_z_block_sign} can vanish.
The pair of final signs specifies the logical Pauli frame according to
\begin{equation}
    (S_X,S_Z)=
    \begin{cases}
        (+1,+1), & I_L,\\
        (+1,-1), & X_L,\\
        (-1,+1), & Z_L,\\
        (-1,-1), & Y_L.
    \end{cases}
    \label{eq:app_qpc_frame_table}
\end{equation}
In particular, an error in the logical $X$-basis sign corresponds to a
logical phase flip, whereas an error in the logical $Z$-basis sign
corresponds to a logical bit flip.

\subsection{Closed-form branch error probabilities}
\label{app:qpc_sector_probabilities}

Under the assumption that the rails are independent and identically
distributed, the branch-error probabilities follow directly from the
binomial distribution.  In the $X$ branch, a block parity is wrong when an
odd number of its $m$ rail signs are wrong.  Hence
\begin{align}
    \pi_X
    &:={\Pr}\bigl[u_i^{(X)}=-1\bigr]\\
    &=\sum_{\substack{k=1\\k\ {\rm odd}}}^{m}
      \binom{m}{k}i_X^k c_X^{m-k}
      =\frac{1-(1-2i_X)^m}{2}.
    \label{eq:app_qpc_x_block_error}
\end{align}
The logical $X$-basis decision is wrong when more than half of the $n$ block
parities are wrong,
\begin{equation}
    E_X
    =\sum_{k=(n+1)/2}^{n}
      \binom{n}{k}\pi_X^k(1-\pi_X)^{n-k}.
    \label{eq:app_qpc_x_error}
\end{equation}

For the $Z$ branch, the error probability after the majority vote inside one
block is
\begin{equation}
    \mu_Z
    :=\sum_{k=(m+1)/2}^{m}
      \binom{m}{k}i_Z^k c_Z^{m-k}.
    \label{eq:app_qpc_z_block_error}
\end{equation}
The product of the $n$ block decisions is wrong precisely when an odd number
of blocks are wrong, giving
\begin{align}
    E_Z
    &=\sum_{\substack{k=1\\k\ {\rm odd}}}^{n}
      \binom{n}{k}\mu_Z^k(1-\mu_Z)^{n-k}\\
    &=\frac{1-(1-2\mu_Z)^n}{2}.
    \label{eq:app_qpc_z_error}
\end{align}
Equations~\eqref{eq:app_qpc_x_block_error}--\eqref{eq:app_qpc_z_error}
are the odd-size, zero-erasure specialization of the QPC decoder derived in
Ref.~\cite{Fukui_2023} (see appendix B4 and B15).  They are also exactly the two branch marginals
evaluated by the numerical implementation.  Here $E_X$ denotes the
probability of an erroneous logical $X$-basis outcome and $E_Z$ the
probability of an erroneous logical $Z$-basis outcome; they are not,
respectively, the probabilities of Pauli $X_L$ and $Z_L$ frames.

For the QPC$(3,3)$ used throughout the repeater comparison, the preceding
expressions reduce to
\begin{align}
    \pi_X&=3i_X-6i_X^2+4i_X^3,
    &
    E_X&=3\pi_X^2-2\pi_X^3,\\
    \mu_Z&=3i_Z^2-2i_Z^3,
    &
    E_Z&=3\mu_Z-6\mu_Z^2+4\mu_Z^3.
    \label{eq:app_qpc_33_decoder}
\end{align}

\subsection{Logical PTM and composition over many hops}
\label{app:qpc_logical_channel}

The analytic pipeline retains $E_X$ and $E_Z$ but not their shot-resolved
joint distribution.  To obtain a four-component logical Pauli channel, it
uses the product-marginal closure
\begin{align}
    p_I^L&=(1-E_X)(1-E_Z),
    &
    p_X^L&=(1-E_X)E_Z,\\
    p_Z^L&=E_X(1-E_Z),
    &
    p_Y^L&=E_XE_Z.
    \label{eq:app_qpc_product_closure}
\end{align}
This is an explicit modeling assumption similar to \cite{Fukui_2023}. A shot-resolved decoder that retains the common rail labels
can instead estimate their joint distribution directly. As a numerical check of the analytic outer-code combinatorics, we additionally sampled $5\times10^{6}$ independent residual rail-level Pauli patterns per code at the zero-propagation-loss threshold point $s=5.059\,\mathrm{dB}$ ($\eta=1$, with no erasure or discard option) and applied the joint single-shot QPC decoder.  For QPC$(5,3)$ and QPC$(13,5)$, the sampled branch marginals agreed with the analytic expressions within $2.79\times10^{-4}$. With the closure in Eq.~\eqref{eq:app_qpc_product_closure}, the one-hop
logical PTM in the ordered basis $(I,X,Y,Z)$ is
\begin{equation}
    C_L^{\varepsilon,\eta}
    =\operatorname{diag}\!\left(
      1,
      1-2E_X,
      (1-2E_X)(1-2E_Z),
      1-2E_Z
      \right).
    \label{eq:app_qpc_logical_ptm}
\end{equation}
For $N$ identical correction hops, the two basis-error probabilities used in
the BB84 calculation are therefore
\begin{align}
    Q_X^{(N)}&=\frac{1-(1-2E_X)^N}{2},
    &
    Q_Z^{(N)}&=\frac{1-(1-2E_Z)^N}{2}.
    \label{eq:app_qpc_chain_qber}
\end{align}
The secret-key fraction per transmitted optical mode is
\begin{equation}
    r_0
    =\frac{1}{nm}
      \max\!\left\{0,
      1-h_2\!\left(Q_X^{(N)}\right)
       -h_2\!\left(Q_Z^{(N)}\right)
      \right\}.
    \label{eq:app_qpc_rate_per_mode}
\end{equation}

\subsection{Relation to the decoder of Fukui et al.}
\label{app:qpc_fukui_relation}

The parity-within-block/majority-across-block rule for the $X$ basis and the
majority-within-block/parity-across-block rule for the $Z$ basis are adopted
from Ref.~\cite{Fukui_2023}.  The input supplied to that outer classical
decoder is different here.  Fukui et al.\ study ideal GKP qubits under an
additive Gaussian displacement channel and use a highly reliable measurement
to map a continuous syndrome to one of three symbols, $\{+1,-1,E\}$, where
$E$ is a located erasure.  In the results of this work, the finite-squeezing
teleportation decoder acts after a pure-loss channel and deterministically
maps each homodyne pair to a Pauli-frame label.  Equivalently, the erasure
probability is set to zero, corresponding to the conventional no-discard
limit of the outer decoder.  We therefore borrow the QPC decision algebra,
but not the highly reliable measurement or the additive-Gaussian physical
model.  No analog likelihood is passed between rails: the continuous
syndrome is used only by the local GKP decoder, after which the outer code
receives hard binary sector data.  For odd $n$ and $m$, the resulting
classical post-processing requires $O(nm)$ operations per shot.

\bibliography{apssamp}

@misc{huang2025vastworldquantumadvantage,
      title={The vast world of quantum advantage}, 
      author={Hsin-Yuan Huang and Soonwon Choi and Jarrod R. McClean and John Preskill},
      year={2025},
      eprint={2508.05720},
      archivePrefix={arXiv},
      primaryClass={quant-ph},
      url={https://arxiv.org/abs/2508.05720}, 
}

@misc{grover1996fastquantummechanicalalgorithm,
      title={A fast quantum mechanical algorithm for database search}, 
      author={Lov K. Grover},
      year={1996},
      eprint={quant-ph/9605043},
      archivePrefix={arXiv},
      primaryClass={quant-ph},
      url={https://arxiv.org/abs/quant-ph/9605043}, 
}

@article{Shor_1997,
   title={Polynomial-Time Algorithms for Prime Factorization and Discrete Logarithms on a Quantum Computer},
   volume={26},
   ISSN={1095-7111},
   url={http://dx.doi.org/10.1137/S0097539795293172},
   DOI={10.1137/s0097539795293172},
   number={5},
   journal={SIAM Journal on Computing},
   publisher={Society for Industrial & Applied Mathematics (SIAM)},
   author={Shor, Peter W.},
   year={1997},
   month=oct, pages={1484–1509} }

@article{bigelRevLett.81.5932,
  title = {Quantum Repeaters: The Role of Imperfect Local Operations in Quantum Communication},
  author = {Briegel, H.-J. and D\"ur, W. and Cirac, J. I. and Zoller, P.},
  journal = {Phys. Rev. Lett.},
  volume = {81},
  issue = {26},
  pages = {5932--5935},
  numpages = {0},
  year = {1998},
  month = {Dec},
  publisher = {American Physical Society},
  doi = {10.1103/PhysRevLett.81.5932},
  url = {https://link.aps.org/doi/10.1103/PhysRevLett.81.5932}
}

@article{Munro2015InsideQR,
  title={Inside Quantum Repeaters},
  author={William J. Munro and Koji Azuma and Kiyoshi Tamaki and Kae Nemoto},
  journal={IEEE Journal of Selected Topics in Quantum Electronics},
  year={2015},
  volume={21},
  pages={78-90},
  url={https://api.semanticscholar.org/CorpusID:33407053}
}

@article{RevModPhys.95.045006,
  title = {Quantum repeaters: From quantum networks to the quantum internet},
  author = {Azuma, Koji and Economou, Sophia E. and Elkouss, David and Hilaire, Paul and Jiang, Liang and Lo, Hoi-Kwong and Tzitrin, Ilan},
  journal = {Rev. Mod. Phys.},
  volume = {95},
  issue = {4},
  pages = {045006},
  numpages = {66},
  year = {2023},
  month = {Dec},
  publisher = {American Physical Society},
  doi = {10.1103/RevModPhys.95.045006},
  url = {https://link.aps.org/doi/10.1103/RevModPhys.95.045006}
}

@article{Muralidharan2015OptimalAF,
  title={Optimal architectures for long distance quantum communication},
  author={Sreraman Muralidharan and Linshu Li and Jungsang Kim and Norbert L{\"u}tkenhaus and Mikhail D. Lukin and Liang Jiang},
  journal={Scientific Reports},
  year={2015},
  volume={6},
  url={https://api.semanticscholar.org/CorpusID:14597684}
}

@article{Gottesman_2001,
   title={Encoding a qubit in an oscillator},
   volume={64},
   ISSN={1094-1622},
   url={http://dx.doi.org/10.1103/PhysRevA.64.012310},
   DOI={10.1103/physreva.64.012310},
   number={1},
   journal={Physical Review A},
   publisher={American Physical Society (APS)},
   author={Gottesman, Daniel and Kitaev, Alexei and Preskill, John},
   year={2001},
   month=jun }

@article{Ralph_2005,
   title={Loss-Tolerant Optical Qubits},
   volume={95},
   ISSN={1079-7114},
   url={http://dx.doi.org/10.1103/PhysRevLett.95.100501},
   DOI={10.1103/physrevlett.95.100501},
   number={10},
   journal={Physical Review Letters},
   publisher={American Physical Society (APS)},
   author={Ralph, T. C. and Hayes, A. J. F. and Gilchrist, Alexei},
   year={2005},
   month=aug }

@article{Fukui_2023,
   title={Efficient Concatenated Bosonic Code for Additive Gaussian Noise},
   volume={131},
   ISSN={1079-7114},
   url={http://dx.doi.org/10.1103/PhysRevLett.131.170603},
   DOI={10.1103/physrevlett.131.170603},
   number={17},
   journal={Physical Review Letters},
   publisher={American Physical Society (APS)},
   author={Fukui, Kosuke and Matsuura, Takaya and Menicucci, Nicolas C.},
   year={2023},
   month=oct }

@article{Muralidharan_2014,
   title={Ultrafast and Fault-Tolerant Quantum Communication across Long Distances},
   volume={112},
   ISSN={1079-7114},
   url={http://dx.doi.org/10.1103/PhysRevLett.112.250501},
   DOI={10.1103/physrevlett.112.250501},
   number={25},
   journal={Physical Review Letters},
   publisher={American Physical Society (APS)},
   author={Muralidharan, Sreraman and Kim, Jungsang and Lütkenhaus, Norbert and Lukin, Mikhail D. and Jiang, Liang},
   year={2014},
   month=jun }

@article{Noh_2019,
   title={Quantum Capacity Bounds of Gaussian Thermal Loss Channels and Achievable Rates With Gottesman-Kitaev-Preskill Codes},
   volume={65},
   ISSN={1557-9654},
   url={http://dx.doi.org/10.1109/TIT.2018.2873764},
   DOI={10.1109/tit.2018.2873764},
   number={4},
   journal={IEEE Transactions on Information Theory},
   publisher={Institute of Electrical and Electronics Engineers (IEEE)},
   author={Noh, Kyungjoo and Albert, Victor V. and Jiang, Liang},
   year={2019},
   month=apr, pages={2563–2582} }

@misc{chatterjee2026allopticalquantummemoryusing,
      title={All-optical quantum memory using bosonic quantum error correction codes}, 
      author={Kaustav Chatterjee and Niklas Budinger and Kian Latifi Yaghin and Lucas Borg Clausen and Ulrik Lund Andersen},
      year={2026},
      eprint={2603.21721},
      archivePrefix={arXiv},
      primaryClass={quant-ph},
      url={https://arxiv.org/abs/2603.21721}, 
}

@article{Pirandola_2017,
   title={Fundamental limits of repeaterless quantum communications},
   volume={8},
   ISSN={2041-1723},
   url={http://dx.doi.org/10.1038/ncomms15043},
   DOI={10.1038/ncomms15043},
   number={1},
   journal={Nature Communications},
   publisher={Springer Science and Business Media LLC},
   author={Pirandola, Stefano and Laurenza, Riccardo and Ottaviani, Carlo and Banchi, Leonardo},
   year={2017},
   month=apr }

@article{PhysRevA.108.052413,
  title = {Analysis of loss correction with the Gottesman-Kitaev-Preskill code},
  author = {Hastrup, Jacob and Andersen, Ulrik Lund},
  journal = {Phys. Rev. A},
  volume = {108},
  issue = {5},
  pages = {052413},
  numpages = {9},
  year = {2023},
  month = {Nov},
  publisher = {American Physical Society},
  doi = {10.1103/PhysRevA.108.052413},
  url = {https://link.aps.org/doi/10.1103/PhysRevA.108.052413}
}

@article{PhysRevA.102.062411,
  title = {Continuous-variable gate teleportation and bosonic-code error correction},
  author = {Walshe, Blayney W. and Baragiola, Ben Q. and Alexander, Rafael N. and Menicucci, Nicolas C.},
  journal = {Phys. Rev. A},
  volume = {102},
  issue = {6},
  pages = {062411},
  numpages = {19},
  year = {2020},
  month = {Dec},
  publisher = {American Physical Society},
  doi = {10.1103/PhysRevA.102.062411},
  url = {https://link.aps.org/doi/10.1103/PhysRevA.102.062411}
}

@article{Nielsen_2002,
   title={A simple formula for the average gate fidelity of a quantum dynamical operation},
   volume={303},
   ISSN={0375-9601},
   url={http://dx.doi.org/10.1016/S0375-9601(02)01272-0},
   DOI={10.1016/s0375-9601(02)01272-0},
   number={4},
   journal={Physics Letters A},
   publisher={Elsevier BV},
   author={Nielsen, Michael A},
   year={2002},
   month=oct, pages={249–252} }

@misc{hler2024quantumrepeatersbasedstationary,
      title={Quantum repeaters based on stationary Gottesman-Kitaev-Preskill qubits}, 
      author={Stefan Häussler and Peter van Loock},
      year={2024},
      eprint={2406.07158},
      archivePrefix={arXiv},
      primaryClass={quant-ph},
      url={https://arxiv.org/abs/2406.07158}, 
}

@article{Bennett_2014,
   title={Quantum cryptography: Public key distribution and coin tossing},
   volume={560},
   ISSN={0304-3975},
   url={http://dx.doi.org/10.1016/j.tcs.2014.05.025},
   DOI={10.1016/j.tcs.2014.05.025},
   journal={Theoretical Computer Science},
   publisher={Elsevier BV},
   author={Bennett, Charles H. and Brassard, Gilles},
   year={2014},
   month=dec, pages={7–11} }

@article{PhysRevA.87.052315,
  title = {Quantum repeaters and quantum key distribution: Analysis of secret-key rates},
  author = {Abruzzo, Silvestre and Bratzik, Sylvia and Bernardes, Nadja K. and Kampermann, Hermann and van Loock, Peter and Bru\ss{}, Dagmar},
  journal = {Phys. Rev. A},
  volume = {87},
  issue = {5},
  pages = {052315},
  numpages = {21},
  year = {2013},
  month = {May},
  publisher = {American Physical Society},
  doi = {10.1103/PhysRevA.87.052315},
  url = {https://link.aps.org/doi/10.1103/PhysRevA.87.052315}
}

@article{Rozp_dek_2021,
   title={Quantum repeaters based on concatenated bosonic and discrete-variable quantum codes},
   volume={7},
   ISSN={2056-6387},
   url={http://dx.doi.org/10.1038/s41534-021-00438-7},
   DOI={10.1038/s41534-021-00438-7},
   number={1},
   journal={npj Quantum Information},
   publisher={Springer Science and Business Media LLC},
   author={Rozpędek, Filip and Noh, Kyungjoo and Xu, Qian and Guha, Saikat and Jiang, Liang},
   year={2021},
   month=jun }

@article{Brady_2024,
   title={Advances in bosonic quantum error correction with Gottesman–Kitaev–Preskill Codes: Theory, engineering and applications},
   volume={93},
   ISSN={0079-6727},
   url={http://dx.doi.org/10.1016/j.pquantelec.2023.100496},
   DOI={10.1016/j.pquantelec.2023.100496},
   journal={Progress in Quantum Electronics},
   publisher={Elsevier BV},
   author={Brady, Anthony J. and Eickbusch, Alec and Singh, Shraddha and Wu, Jing and Zhuang, Quntao},
   year={2024},
   month=Jan, pages={100496} }

@article{Tzitrin_2020,
   title={Progress towards practical qubit computation using approximate Gottesman-Kitaev-Preskill codes},
   volume={101},
   ISSN={2469-9934},
   url={http://dx.doi.org/10.1103/PhysRevA.101.032315},
   DOI={10.1103/physreva.101.032315},
   number={3},
   journal={Physical Review A},
   publisher={American Physical Society (APS)},
   author={Tzitrin, Ilan and Bourassa, J. Eli and Menicucci, Nicolas C. and Sabapathy, Krishna Kumar},
   year={2020},
   month=Mar }

@article{Fukui_2018,
   title={High-Threshold Fault-Tolerant Quantum Computation with Analog Quantum Error Correction},
   volume={8},
   ISSN={2160-3308},
   url={http://dx.doi.org/10.1103/PhysRevX.8.021054},
   DOI={10.1103/physrevx.8.021054},
   number={2},
   journal={Physical Review X},
   publisher={American Physical Society (APS)},
   author={Fukui, Kosuke and Tomita, Akihisa and Okamoto, Atsushi and Fujii, Keisuke},
   year={2018},
   month=May }

@article{Fukui_2017,
   title={Analog Quantum Error Correction with Encoding a Qubit into an Oscillator},
   volume={119},
   ISSN={1079-7114},
   url={http://dx.doi.org/10.1103/PhysRevLett.119.180507},
   DOI={10.1103/physrevlett.119.180507},
   number={18},
   journal={Physical Review Letters},
   publisher={American Physical Society (APS)},
   author={Fukui, Kosuke and Tomita, Akihisa and Okamoto, Atsushi},
   year={2017},
   month=Nov }

@article{Noh_2022,
   title={Low-Overhead Fault-Tolerant Quantum Error Correction with the Surface-GKP Code},
   volume={3},
   ISSN={2691-3399},
   url={http://dx.doi.org/10.1103/PRXQuantum.3.010315},
   DOI={10.1103/prxquantum.3.010315},
   number={1},
   journal={PRX Quantum},
   publisher={American Physical Society (APS)},
   author={Noh, Kyungjoo and Chamberland, Christopher and Brandão, Fernando G.S.L.},
   year={2022},
   month=Jan }
\end{document}